\documentclass[aps,twocolumn,prd,longbibliography,superscriptaddress,preprintnumbers,
nofootinbib,showpacs]{revtex4-2}

\usepackage[utf8]{inputenc}
\usepackage{graphicx}
\graphicspath{{./figures/}}
\usepackage{dcolumn}
\usepackage{amsmath}
\usepackage{amssymb}
\usepackage{float}
\usepackage{vmargin}
\usepackage[dvipsnames]{xcolor} 
\definecolor{lcolor}{rgb}{0.5,0,0}
\definecolor{citcolor}{rgb}{0,0,1}
\usepackage[breaklinks,colorlinks,urlcolor=blue,citecolor=citcolor,linkcolor=lcolor,linktoc=all]{hyperref}
\usepackage{orcidlink}

\newcommand{\chiinv}[2]{\left(\chi^{-1}\right)_{#1}^{\ #2}}
\newcommand{\chiinvdiag}[2]{\chi_{#1#2}^{-1}}
\newcommand{\conv}[1]{D_#1}
\newcommand{\Tmatrix}{\mathcal{T}}
\newcommand{\heaviside}{\theta_\mathrm{H}}

\allowdisplaybreaks

\setpapersize{A4}
\setmargins{2.5cm}       
{1.5cm}                    
{17cm}                   
{23.42cm}              
{10pt}                 
{1cm}              
{0pt}                
{2cm}

\begin{document}

\title{Burgers equation for the bulk viscous pressure of  quark matter}
\author{José Luis Hernández
\orcidlink{0000-0002-7241-2843}}
\email{hernandez@ice.csic.es}
\affiliation{Institute of Space Sciences (ICE, CSIC), Campus UAB,  Carrer de Can Magrans, 08193 Barcelona, Spain}
\affiliation{Institut d'Estudis Espacials de Catalunya (IEEC), 08034 Barcelona, Spain}
\affiliation{Facultat de Física, Universitat de Barcelona, Martí i Franquès 1, 08028 Barcelona, Spain.}
\author{Cristina Manuel
\orcidlink{0000-0002-0024-3366}}
\email{cristina.manuel@csic.es}
\affiliation{Institute of Space Sciences (ICE, CSIC), Campus UAB,  Carrer de Can Magrans, 08193 Barcelona, Spain}
\affiliation{Institut d'Estudis Espacials de Catalunya (IEEC), 08034 Barcelona, Spain}
\author{Saga Säppi
\orcidlink{0000-0002-2920-8038}}
\email{sappi@ieec.cat}
\affiliation{Institute of Space Sciences (ICE, CSIC), Campus UAB,  Carrer de Can Magrans, 08193 Barcelona, Spain}
\affiliation{Institut d'Estudis Espacials de Catalunya (IEEC), 08034 Barcelona, Spain}
\author{Laura Tolos
\orcidlink{0000-0003-2304-7496}}
\email{tolos@ice.csic.es} 
\affiliation{Institute of Space Sciences (ICE, CSIC), Campus UAB,  Carrer de Can Magrans, 08193 Barcelona, Spain}
\affiliation{Institut d'Estudis Espacials de Catalunya (IEEC), 08034 Barcelona, Spain}

\date{June 2025}

\begin{abstract}

The dissipative properties of relativistic strongly interacting nuclear 
matter significantly influence the damping of stellar oscillations and 
density fluctuations during compact star mergers. In this work, we 
derive the evolution equation for the bulk viscous pressure in unpaired 
quark matter under small deviations from equilibrium. Our analysis 
reveals that it behaves like a two-component Burgers fluid. We identify 
four key transport coefficients—two relaxation times and two bulk 
viscosity coefficients—expressed in terms of equilibrium parameters and electroweak nonleptonic and semi-leptonic decay rates.
The transport coefficients are evaluated for two distinct equations of state: one based on 
perturbative quantum chromodynamics and the other on a modified MIT bag 
model, valid in different density regimes. We also determine the temperature and density region where nonleptonic  electroweak processes dominate the dissipation.  Our formulation establishes a new way of describing bulk viscous effects in quark matter, applicable for numerical simulations of compact star mergers.

\end{abstract}


\maketitle

\section{Introduction}

Quantum Chromodynamics (QCD), the fundamental theory describing strong nuclear interactions, predicts the existence of exotic phases of matter under extreme conditions \cite{qcdbook,WEBER2005193,Alford_2008}. In particular, at the high densities anticipated in the cores of compact stars quarks may become deconfined. There are many methods of detecting these  novel forms of matter \cite{Alford:2019oge}. While the equation of state (EOS) of dense matter constrains the mass-to-radius ratio of compact objects, out-of-equilibrium phenomena may give the key to understanding their dynamical behavior. 
With recent and future advances in gravitational-wave astronomy \cite{LIGOScientific:2017vwq,LIGOScientific:2017ync,Faber:2012rw,Baiotti:2016qnr}, both the EOS and the transport properties of dense matter will be constrained more stringently beyond what is possible through electromagnetic observations alone.
It is then important to deepen our understanding of the distinctive signatures of the quark matter phases in these compact star scenarios, and also provide suitable methods for their study.

In this Article we focus on the bulk viscous effects of unpaired quark matter, that is, with the dissipation arising from compression and rarefaction. These effects play a crucial role in damping the rapid oscillations that follow the merger of compact stars \cite{Most:2022yhe,Chabanov:2023blf,Chabanov:2023abq}, as well as various stellar oscillation modes in isolated stars \cite{Kokkotas:1999bd,Rezzolla:2003ua}, both of which could influence gravitational wave signals \cite{Ghosh:2025wfx}. Previous studies have computed bulk viscosity coefficients for different hadronic
\cite{Sawyer:1989dp,Haensel:1992zz,Haensel:2000vz,Alford:2018lhf,Alford:2019qtm,Alford:2023gxq} and quark matter phases \cite{Sawyer:1989uy,Madsen:1992sx,Sad:2007afd,CruzRojas:2024etx,Hernandez:2024rxi}  (see \cite{Harris:2024evy} for a review and a more complete set of references), typically as functions of the oscillation frequency, allowing quick estimates of energy dissipation. However, such formulations are not readily implementable in numerical simulations of the relativistic hydrodynamics associated with the relevant astrophysical settings.
Only recently have efforts begun to incorporate bulk viscous dissipation into simulations of neutron star mergers, typically assuming that the bulk viscous stress tensor satisfies an Israel--Stewart (IS) equation \cite{Israel:1979wp,Hiscock:1983zz,Bemfica:2019cop,Yang:2023ogo,Yang:2024dsv,Yang:2023ozd,Yang:2025yoo,Harutyunyan:2023nvt}. This second-order hydrodynamic framework introduces a relaxation time, an additional transport coefficient, in addition to the bulk viscosity itself, ensuring the causal propagation of hydrodynamical perturbations, in contrast to standard first-order formulations.

In this work we show that the bulk viscous pressure in unpaired quark matter evolves according to a Burgers-type equation, also used in viscoelastic media \cite{viscobook}.  In \cite{Gavassino:2023eoz} Gavassino has shown that the bulk pressure of systems with three independent chemical potentials and two distinct reaction rates, such as hadronic matter composed of neutrons, protons, electrons and muons, also follows Burgers-type evolution. While quark matter does not fulfill the above mentioned conditions, in this Article we show that its bulk viscous pressure is also governed by a Burgers equation. Only in certain temperature and density regimes does it reduce to the more familiar IS form. We compute the full set of second-order transport coefficients entering the Burgers equation, conveniently expressed in terms of two partial bulk viscosity coefficients and two relaxation times. 
Although our formulation is general, we explicitly evaluate these transport coefficients for two distinct quark matter EOSs at finite densities, temperatures, and for a finite strange quark mass: one derived from perturbative QCD (pQCD), applicable at very high densities \cite{Gorda:2021gha,Gorda:2023mkk,Kurkela:2016was,Kurkela:2009gj,Fraga:2004gz,Laine:2006cp}, and another based on a modified MIT bag model \cite{Torres_2013,Lopes_2021}, more suitable for lower densities. Our approach provides a versatile framework that can accommodate more accurate EOS models or updated reaction rates as they become available, and can provide the proper description of the limiting IS case as well. 

We use metric conventions $(+,-,-,-)$ and natural units $k_B= \hbar=c=1$ throughout, unless otherwise stated.

\section{Setup}

 For definiteness let us consider a situation relevant to the astrophysical settings. We take into account the presence of five species of
particles:  massless $u,d$-quarks, massive $s$-quarks, massless electrons $e$, each with a chemical potential $\mu_{i}$, as well as neutrinos $\nu$. After a disturbance, the chemical
(beta) equilibrium is achieved by $W$-boson-mediated electroweak processes,
namely 
\begin{eqnarray}
\label{eq:eqlprocesses}
&& u+d \longleftrightarrow u+s , \nonumber \\
&& u+e \longrightarrow d+\nu , \quad  
 d \longrightarrow u+e+\bar{\nu} , \nonumber \\
&& u+e \longrightarrow s+\nu , \quad
 s \longrightarrow u+e+\bar{\nu} .
\end{eqnarray}
Of these, the first one is the only \emph{nonleptonic} process, whereas the remaining four are \emph{semileptonic}. 
We assume that neutrinos remain untrapped \cite{Iwamoto:1982zz}, restricting to temperatures below $10~\,\mathrm{MeV}$, ignoring neutrino chemical
potentials. The flavor-changing processes above lead to two linearly
independent chemical potentials which characterize deviation from
equilibrium. We define
\begin{equation}
\mu_{1}\equiv\mu_{s}-\mu_{d},  \qquad 
\quad\mu_{2}\equiv\mu_{u}-\mu_{d}+\mu_{e},
\label{eq:mu12def}
\end{equation}
noting that the chemical potential that characterizes the reaction
$u+e\longleftrightarrow s+\nu$ is $\mu_{u}-\mu_{s}+\mu_{e}=\mu_{2}-\mu_{1}$. In terms of these two chemical potentials, we can write the linearized reaction rates
\begin{align}
    \Gamma_{s+u \rightarrow d+u} - \Gamma_{d+u \rightarrow s+u}&\equiv\mu_1 \lambda_1 , \nonumber \\
    \Gamma_{d \rightarrow u+e+\bar{\nu}_e} - \Gamma_{u+e \rightarrow d+\nu_e}&\equiv-\mu_2 \lambda_2, \nonumber \\
    \Gamma_{s \rightarrow u+e+\bar{\nu}_e} - \Gamma_{u+e \rightarrow s+\nu_e}&\equiv(\mu_2-\mu_1)\lambda_3.
\end{align}
The transport properties will also depend on (the partial derivatives of) the equilibrium pressure $p$, specifically on the particle number densities $n_a$ and susceptibilities $\chi_{ab}$ defined as
\begin{equation}
    n_a \equiv \frac{\partial p}{\partial \mu_a},\,\, \chi_{ab} \equiv \frac{\partial n_a}{\partial \mu_b} = \frac{\partial^2 p}{\partial \mu_a \partial \mu _b},\,\, a,b\in\lbrace u,d,s,e \rbrace. 
\end{equation}
As we will linearize our equations in $\mu_1,\mu_2$, susceptibilities and densities are always taken to be evaluated in equilibrium, with the chemical potentials subject to two additional beta-equilibrium constraints, $\mu_{s}=\mu_{d}=\mu_{u}+\mu_{e}$. In addition to constraints arising from the reactions, the chemical potentials have to fullfill the requirement of charge neutrality, which is most conveniently written down as
$n_{Q}\equiv\left(2n_{u}-n_{d}-n_{s}\right)/3-n_{e}=0.$

\section{Burgers equation}

In order to study the out-of-equilibrium evolution of quark matter, we write down equations for the bulk scalar
$\Pi =p_{\mathrm{non-eq.}}-p$, measuring the difference of the 
 non-equilibrium and equilibrium pressures, as well as the fluid velocity $u^\mu$ and
 $\vartheta \equiv \partial_\mu u^\mu$. In the system we consider
\begin{align}
\Pi&= n_u \left( \delta\mu_u + \delta \mu_e - \delta\mu_s \right) \nonumber\\
&+ n_d \left(\delta\mu_d-\delta\mu_s\right) 
+ n_q (\delta\mu_s-\frac{1}{3}\delta\mu_e), 
\label{eq:bulkscalardefn}
\end{align}
where $\delta \mu_a$ measures the deviation of the chemical potential of the particle $a$ with respect to its equilibrium value, and $n_{q}\equiv n_{u}+n_{d}+n_{s}$
is the quark density. Constraints of charge neutrality and conservation of baryon number have to be imposed (see Eqs.~(\ref{eq:charge-constraint}, \ref{eq:baryon-constraint}) of Appendix \ref{sec:constraints}) to express $\Pi$ in terms of the algebraically independent chemical potentials $\mu_1,\mu_2$. This amounts to writing the chemical potentials appearing in Eq. \eqref{eq:bulkscalardefn} in terms of $\mu_1,\mu_2$, densities $n_a$, and (inverse) susceptibilities $\chi_{ab}^{(-1)}$. As a consequence of the constraints, the equilibrium system is characterized by a single free chemical potential which we choose to be $\mu_d$.

We follow the logic of \cite{Gavassino:2023eoz},
adapting the notation and derivation for our purposes. To obtain a differential equation for the chemical potentials, we generalize the equations found in the Supplemental Material of \cite{CruzRojas:2024etx}  to non-oscillatory
deformations of the fluid. That is, we start from the evolution equations of the different particle fractions, and
after linearizing them around  equilibrium,
we find that the chemical potentials fulfill the system of differential
equations
\begin{equation}
\conv{u}\mu_{\Gamma}=\left[M\lambda\hat{\boldsymbol{\mu}}\right]_{\Gamma}-\left[M\boldsymbol{X}\right]_{\Gamma}n_{q}\vartheta,\, \Gamma\in\left\{ 1,2,s,e\right\} ,
\end{equation}
where $\conv{u}= u_\mu \partial^\mu$ is the convective derivative, and 
the indices $\{u,d,s,e\}$ refer to the particle species, whereas
$\{1,2\}$ refer to combinations corresponding to $\mu_{1},\mu_{2}$
defined above. The matrix $M$ contains  elements of the inverse susceptibility matrix, and $\lambda$ is a matrix of the rates of the weak reactions,
explicitly
\begin{align}
\left(M\right)_{\Gamma}^{\,\,a} &\equiv \chiinv{\Gamma}{a},\\
\chiinv{1}{a}&\equiv \chiinv{s}{a} - \chiinv{d}{a}, \\
\chiinv{2}{a}&\equiv \chiinv{u}{a} + \chiinv{e}{a} - \chiinv{d}{a}; \\
\lambda&\equiv\begin{pmatrix}\lambda_{3} & -\lambda_{2}-\lambda_{3}\\
\lambda_{1} & \lambda_{2}\\
-\lambda_{1}-\lambda_{3} & \lambda_{3}\\
\lambda_{3} & -\lambda_{2}-\lambda_{3}
\end{pmatrix}.
\end{align}
Lastly, $\boldsymbol{X}\equiv\left(n_{u},n_{d},n_q-n_{u}-n_{d},n_{u}-n_q/3\right)/n_{q}$ 
is a vector of the (equilibrium) particle fractions and $\hat{\boldsymbol{\mu}}~\equiv~\left(\mu_{1},\mu_{2}\right)$.  Using the above quantities as well as $\hat{M}$, the reduction of $M$ into
its first two rows, we now define a two-by-two matrix $\Tmatrix$ \footnote{The matrix is well-defined for reasonable EOSs as
long as the reaction rates are all nonzero, this is discussed in more
detail in \cite{Gavassino:2023eoz}.}, and a two-component vector $\boldsymbol{b}$: 
\begin{equation}
    \Tmatrix^{-1} \equiv - \hat{M}\lambda,\quad \boldsymbol{b}\equiv n_q\Tmatrix\hat{M}\boldsymbol{X}. 
\end{equation}
We now readily find that
\begin{equation}
\left(\mathrm{id}_2+\Tmatrix\conv{u}\right)\hat{\boldsymbol{\mu}}=-\vartheta\boldsymbol{b}.
\end{equation}
Here, $\mathrm{id}_2$ is the two-by-two identity matrix.  By taking the second convective derivative of this equation, noting that the convective derivatives of equilibrium quantities vanish, multiplying the equation by $\det\left(\Tmatrix\right)\Tmatrix^{-1}$, and using the identity $\mathrm{Tr}\left(\Tmatrix\right)\mathrm{id}_{2}-\det\left(\Tmatrix\right)\Tmatrix^{-1}=\Tmatrix$,
valid for any non-singular two-by-two matrix,
we obtain 
\begin{widetext}
\begin{equation}
\label{eq:burgers}
\det\left(\Tmatrix\right)\conv{u}^{2}\hat{\boldsymbol{\mu}}+\mathrm{Tr}\left(\Tmatrix\right)\conv{u}\hat{\boldsymbol{\mu}}+\hat{\boldsymbol{\mu}}=-\vartheta\boldsymbol{b}-\left(\conv{u}\vartheta\right)\det \left( \Tmatrix\right)\Tmatrix^{-1}\boldsymbol{b}.
\end{equation}
\end{widetext}

If we now write $\tau_{\pm}$ for the eigenvalues of $\Tmatrix$,
define $\left(\Pi_{1},\Pi_{2}\right)\equiv\boldsymbol{\Pi}$ via $\Pi=\Pi_{1}\mu_{1}+\Pi_{2}\mu_{2}$,
and contract \eqref{eq:burgers} with it, we get the Burgers equation valid for our system: 
\begin{equation}
\label{Burgers-quark}
\tau_{+}\tau_{-}\conv{u}^{2}\Pi +\left(\tau_{+}+\tau_{-}\right)\conv{u}\Pi+\Pi=-\zeta\vartheta-\xi \conv{u}\vartheta ,
\end{equation}
with transport coefficients 
\begin{align}
\zeta&=\left\langle \boldsymbol{\Pi},\boldsymbol{b}\right\rangle 
,\qquad \xi= \tau_{+}\tau_{-}\left\langle 
\boldsymbol{\Pi},\Tmatrix^{-1}\boldsymbol{b}\right\rangle ,
\label{eq:xidef}
\end{align}
where $\left\langle -,- \right\rangle$ denotes the  (here, two-dimensional) scalar product.
In Eqs. \eqref{eq:xidef} the coefficients $\Pi^{\alpha}$ can be explicitly solved after taking into account the constraints of electrical neutrality and baryon number conservation. The explicit but rather cumbersome forms of the bulk scalar and the transport coefficients can be found in Appendix \ref{sec:explitiformulae}.

The four quantities $\{ \tau_+,\tau_-, \zeta, \xi \}$ characterize the system. However, their interplay
is somewhat subtle. A more quantitative account can be obtained
by examining the corresponding Green's function. As \cite{Gavassino:2023eoz}
points out, the Green's function of the Burgers equation in the fluid rest frame reads, as a function of time $t$,
%
\begin{equation}
G\left(t\right)=G_+\left(t\right)+ G_-\left(t\right) ,  \quad G_{\pm} (t) \equiv \frac{\zeta_{\pm}}{\tau_{\pm}}\heaviside\left(t\right)e^{-t/\tau_{\pm}},
\label{eq:green}
\end{equation}
where $\theta_H$ is the step function, and we have defined the partial viscosities
\begin{equation}
\zeta_{+}=\frac{\zeta\tau_{+}-\xi}{\tau_{+}-\tau_{-}},\qquad\zeta_{-}=\frac{\zeta\tau_{-}-\xi}{\tau_{-}-\tau_{+}}.
\label{eq:zetapm}
\end{equation}
We will find more useful to use $\{\tau_+,\zeta_+,\tau_-,\zeta_- \}$.
Further, while the above derivation is not formally valid when $\mathcal{T}$ is singular, in the limit where one of the relaxation times is very large, $\tau_\pm \rightarrow \infty$, one finds that the hydrodynamical equation reduces to
$\tau_\mp \conv{u} \Pi +\Pi = - \zeta_\mp\vartheta$ and the IS limit is recovered. In particular, this occurs if one neglects the semileptonic processes, as when $\lambda_2,\lambda_3 \rightarrow0$, then $\tau_- \rightarrow \infty$. Otherwise, all three rates enter into the expressions of the two relaxation times and viscosities.

\section{Results}

We apply the formulae derived above to two specific EOSs, with different applicabilities. We consider a modified bag model at intermediate densities ($\mu_d \approx 400\text{--}600~\,\mathrm{MeV}$) and perturbative QCD (pQCD) at high densities ($\mu_d \gtrsim 850~\,\mathrm{MeV}$), both at finite but small temperatures. For reference, for the range of temperatures we cover, below $T\lesssim 10~\,\mathrm{MeV}$, the lower bound of the applicability range  corresponds approximately to $n_B \sim 3.6\, n_\mathrm{sat}$ and $n_B \sim 37\,n_\mathrm{sat}\text{--}38 \,n_\mathrm{sat}$, where $n_\mathrm{sat}\approx 0.16~\,\mathrm{fm}^{-3}$ is the value of the nuclear saturation density, for the bag model and pQCD, respectively. 

In addition to fixing the strong dynamics, we must fix the electroweak rates $\lambda_i$. These are presently known only in a leading-order expansion in the limit of small temperature and strange quark mass (i.e., with the quark chemical potentials as the dominant scales). Making such an assumption, we have \cite{Madsen:1993xx,Heiselberg_1992,Heiselberg:1991px,Koch:1991qh,Iwamoto:1980eb,Iwamoto:1982zz,Anand2009,schwenzer2012longrange} 
\begin{eqnarray}    
    \lambda_1&\approx& \frac{64}{5\pi^3}G_{\mathrm{F}}^2 \sin^2 \Theta_\mathrm{C} \cos^2 \Theta_\mathrm{C} \mu_d^5 T^2, \\
    \lambda_2&\approx&\frac{17}{15 \pi^2} G_{\mathrm{F}}^2 \cos^2 \Theta_\mathrm{C} \alpha_s \mu_d \mu_u \mu_e T^4, \\
    \lambda_3&\approx&\frac{17}{40 \pi} G_{\mathrm{F}}^2 \sin^2 \Theta_\mathrm{C} \mu_s m_s^2 T^4, 
\end{eqnarray}
where $G_{\mathrm{F}}\approx1.166\times 10^{-5}\, \text{GeV}^{-2}$ is the Fermi coupling constant, $\Theta_\mathrm{C}\approx13.02^\circ$ is the Cabibbo angle, $\alpha_s$ is the strong fine-structure constant,  
and $m_s$ is the strange quark mass, $m_s\left(\Lambda=2\,\mathrm{GeV}\right)\approx94\,\mathrm{MeV}$ \cite{ParticleDataGroup:2024cfk}, evaluated at a fixed renormalization scale $\Lambda$ and evolved according to the renormalization group equations. Notably, the only rate scaling with a fifth power of the quark chemical potentials is $\lambda_1$, with the semileptonic processes remaining formally subdominant --- nevertheless, we do not neglect them as they turn out to be crucial for understanding the bulk viscosity $\zeta_-$. We emphasize that in the expression for the rates the values for $m_s$ and $\alpha_s$ must be replaced by their model counterparts in the bag model. In the figures shown in the main text, we have always fixed $m_s$ to its bare value and set the bag model coupling $a_4 \approx 1- 2\alpha_s/\pi$ \cite{Fraga:2001id}  to $a_4=0.7$. We discuss the parameter dependence in more detail in Appendix \ref{sec:constraints}. 

\begin{figure}
\centering
    \includegraphics[width=0.48\textwidth]{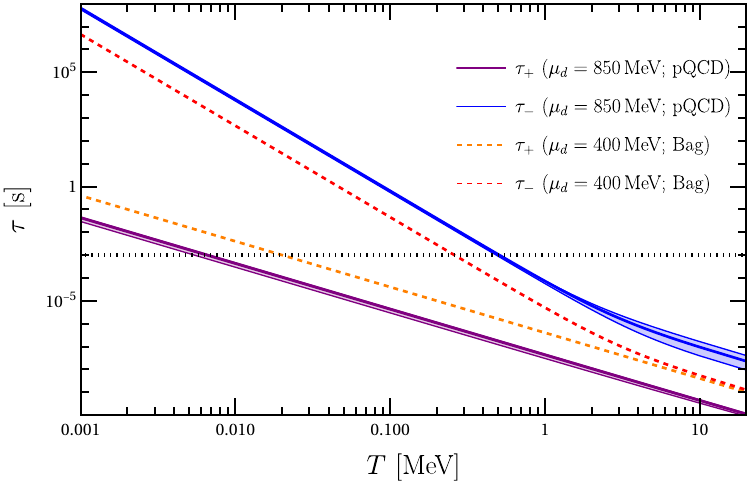}
    \caption{The relaxation times $\tau_\pm$ in seconds, for the bag model and pQCD at $\mu_d = 400$ and $850~\,\mathrm{MeV}$ respectively. We have marked the millisecond scale, relevant for mergers or stellar oscillation modes, and shown the uncertainty bands obtained by varying the renormalization scale in pQCD.}
    \label{fig:tauplot}
\end{figure}

The predictions we obtain appear qualitatively unified, with a caveat for the coefficient $\tau_-$ obtained in the bag model at high densities. In Fig. \ref{fig:tauplot}, we see the relaxation times $\tau_\pm$. For both EOSs $\tau_{+}$ is linear in $ T$ on a log--log-scale, while $\tau_{-}$ is piecewise linear with an inflection point tempering its slope at an $O(\mathrm{MeV})$-scale temperature. There is a qualitative difference here present between the bag model and pQCD, with the pQCD values for $\tau_{-}$ saturating to a nearly density-independent form above $\tau_{+}$. Meanwhile, the $\tau_{-}$ remains strongly density-dependent and attains much smaller values in the bag model, becoming nearly equal to $\tau_{+}$ past the inflection point. We believe that this is in large part due to the form of the rates, where the (constant) effective coupling is used for the bag model, while the standard running coupling is used for pQCD. Thus, at larger temperatures, particularly in combination with relatively large densities, the coupling-dependent $\lambda_2$ is overestimated in the bag model.

\begin{figure}
\centering
    \includegraphics[width=0.48\textwidth]{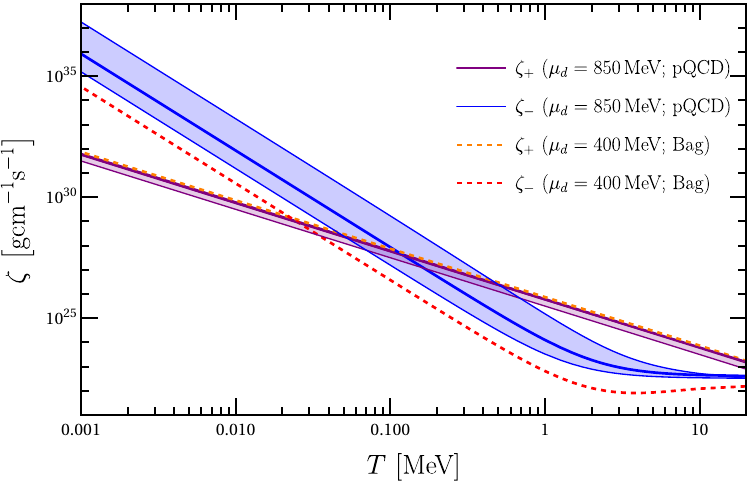}
    \caption{The bulk viscosity components $\zeta_\pm$ for the bag model and pQCD at $\mu_d = 400$ and $850~\,\mathrm{MeV}$ respectively. Values of the bulk viscosity are given in CGS units. The uncertainty bands describe the dependence on the renormalization scale in pQCD.}
    \label{fig:zetaplot}
\end{figure}

The bulk viscosities $\zeta_{\pm}$ behave similarly (see Fig.~\ref{fig:zetaplot}), but $\zeta_-$ saturates to a nearly constant value, and starting at approximately $T=0.1~\,\mathrm{MeV}$, $\zeta_{+}$ overtakes $\zeta_{-}$. The pQCD error bands, obtained by varying the renormalization scale (see Appendix \ref{sec:constraints} for details), are sizable for $\zeta_{-}$ before the inflection point, and considerably more modest for all other quantities. 

The Green's functions encode the effects of the transport coefficients in nontrivial ways. In practice, at a fixed density and time $t'$, the system will start with $G_+(t')$ dominating over $G_-(t')$, and as the fluid heats up, the picture rapidly flips, with $G_-(t)$ crossing over $G_+(t')$ and becoming the dominant contribution. We denote the temperature where this happens, ie, where $G_+(t')=G_-(t')$, as $T_\mathrm{cross}$. At extremely high temperatures, beyond those considered here, neutrinos are trapped, and the behavior of the system is further complicated.

Furthermore, we find that $G_+$ is well-approximated by the nonleptonic limit of $\lambda_2,\lambda_3\ll\lambda_1$, but $G_-$ admits no simple approximation. Thus, we find that bulk viscous effects in quark matter require taking into account both channels above $T_\mathrm{cross}$, and are well-approximated by a single-component fluid, ignoring the semileptonic processes, below $T_\mathrm{cross}$, with a narrow intermediate region.

\begin{figure}
\centering
    \includegraphics[width=0.48\textwidth]{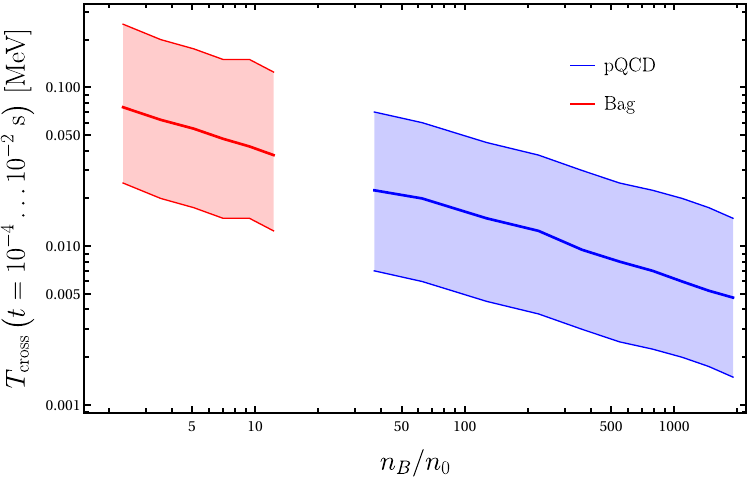}
    \caption{The approximate crossing temperature $T_\mathrm{cross}$ at which the two Green's functions $G_\alpha$ coincide as a function of baryon density $n_B$. Only the central value of the renormalization scale is shown for pQCD. The band represents the time scales $t=10^{-4}\ldots 10 ^{-2}\,\mathrm{s}$, with the thick line corresponding to the millisecond scale. For $T \ll T_\mathrm{cross}$ dissipation is dominated by the nonleptonic processes. }  \label{fig:crossingplot}
\end{figure}

We also note that these temperatures are of relevance for physical neutron star mergers. In merger simulations, the relevant timescales are of millisecond order \cite{Chabanov:2023blf,Chabanov:2023abq}. In  Fig.~\ref{fig:crossingplot}, we show the crossing temperature as a function of the baryon density\footnote{To be specific, as we have finite resolution due to the numerics required to solve $\mu_e$, we find for each density the temperature minimizing $1-G_{+}(t)/G_{-}(t).$}, with a band whose width is determined by the relevant timescales, here taken to be from $10^{-4}\,\mathrm{s}$ to $10^{-2}\,\mathrm{s}$ for illustrative purposes. The region below the band corresponds to the nonleptonic regime dominated by $G_+$, the region above it corresponds to the  regime dominated by $G_-$, and the band to the region where either or both components $G_\alpha$ can be relevant.

\section{Discussion}
Simple relativistic hydrodynamic formulations of viscous effects suffer from significant shortcomings, including acausal signal propagation and inherent instabilities.  These issues were addressed long ago through the development of a more advanced framework, incorporating second-order gradients of hydrodynamic variables. In second-order hydrodynamics, the dissipative component of the stress-energy tensor is elevated to the status of an independent dynamical variable,  alleviating the problems of first-order hydrodynamics. In this work, we have derived an evolution equation for the bulk viscous pressure in unpaired quark matter, to be incorporated in second-order hydrodynamical formulations. 
While bulk viscous dissipation in the simulation of mergers of neutron stars has been studied with an IS evolution equation \cite{Chabanov:2023abq,Chabanov:2023blf}, we have shown that unpaired quark matter is more accurately described with a Burgers equation. It would be very interesting to extend our formulation to other quark matter phases  \cite{Alford:2024tyj,Alford:2025tbp,Mannarelli:2009ia,Bierkandt:2011zp}.

After using two different EOSs, valid for either very large or moderate densities, we delineate the temperature and density regimes where the nonleptonic electroweak processes dominate the dissipation, in time scales that may be relevant for mergers, and could influence the damping of oscillation modes in the star. 
Let us recall here that in these events the effect associated to the shear viscosity \cite{Heiselberg:1993cr}, dominated by the quark-quark scattering  mediated by one-gluon exchange, 
is negligible \cite{Ghosh:2025wfx} in the temperature regime we considered.

Finally, let us comment that while the Burgers equation (\ref{Burgers-quark}) describes the close-to-chemical-equilibrium evolution of quark matter, the same effects could be described by taking into account the individual evolution of the densities of the quarks and electrons, after imposing the constraints of charge neutrality and baryon number conservation in their evolution. However, our proposed formulation is considerably simpler and  will be of great use in analytical models and could facilitate numerical studies of both stellar oscillation modes and mergers of quark or hybrid stars. In future, the treatment found here could also be further extended by considering the variety of superconducting phases found in dense matter.

\section{Acknowledgements}
We  acknowledge support from the program Unidad de Excelencia María de Maeztu CEX2020-001058-M, from the
project PID2022-139427NB-I00 financed by the Spanish MCIN/AEI/10.13039/501100011033/FEDER, UE (FSE+), as well as from the Generalitat de Catalunya under contract 2021 SGR 171. L.T. was also supported by the Grant CIPROM 2023/59 of Generalitat Valenciana.

\section{Data Availability}

The data that support the findings of this article are openly available \cite{data}.

\appendix

\section{Frequency-dependent bulk viscosity }
\label{sec:freqdep}

 From the Burgers equation, one can recover the same value of the frequency-dependent bulk viscosities as found using first-order hydrodynamics. One simply has to assume that the time dependence of the bulk viscous pressure is harmonic $ \Pi \propto e^{i \omega t}$. After solving the Burgers equation, and computing the bulk  viscosity as $ \zeta_{\rm eff} (\omega) = - {\rm Re} \Pi/\vartheta$, one finds
  \begin{equation}
 \label{frequencyzeta}
\zeta_{\rm eff} (\omega)= \frac{ \kappa_1 + \kappa_2 \omega^2}{\kappa_3 + \kappa_4 \omega^2 + \omega^4} ,
 \end{equation}
with the following identifications, 
\begin{align}
\kappa_3 &= \frac{1}{(\tau_+ \tau_-)^2}, \qquad \frac{\kappa_1}{\kappa_3} = \zeta_+ +\zeta_-,\nonumber\\ \qquad  \frac{\kappa_4}{\kappa_3} &= \tau_+^2 + \tau^2_-,  \qquad\,\,\frac{\kappa_2}{\kappa_3} =  \zeta_+\tau_-^2+\zeta_-\tau_+^2.
\end{align}
We have checked that with the values of the   second-order coefficients found in this manuscript we reproduce the frequency-dependent bulk viscosity computed in \cite{Hernandez:2024rxi}\footnote{Note that in \cite{Hernandez:2024rxi}, the chemical potential $\mu_2$ is denoted $-\mu_3$, and the chemical potential $\mu_2$ found there corresponds to our linear combination $\mu_2-\mu_1$. The rates $\lambda_{2,3}$ are likewise swapped.}.
One can even use the above relations to obtain the associated Burgers equation fulfilled by any system that has a frequency-dependent  bulk viscosity
with the same dependence as in Eq.~(\ref{frequencyzeta}).

 When $\tau_- \rightarrow \infty $, the frequency-dependent viscosity reduces to
\begin{equation}
    \zeta_{\rm eff} (\omega) \rightarrow \frac{\zeta_+}{\omega^2 + \tau^2_+}.
\end{equation}
This frequency-dependent bulk viscosity can be recovered from the IS equation  $\tau_+ \conv{u} \Pi + \Pi = -\zeta_+ \vartheta$.

\section{Details of constraints and the equations of state}
\label{sec:constraints}

As we explain in the main text, linearizing quantities in deviations of chemical potentials $\delta \mu_a$ from their equilibrium values leads to formulae dependent only on \emph{equilibrium} susceptibilities and densities. To evaluate them, we enforce beta-equilibrium constraints of $\mu_s=\mu_d=\mu_u+\mu_e$ as well as a charge neutrality constraint $n_Q=(2n_u-n_d-n_s)/3-n_e=0$. These three constraints leave us with only a single free equilibrium chemical potential used to characterize quark matter close to equilibrium. In practice, we choose this to be the $d$-quark chemical potential $\mu_d$, and solve the (small) electron chemical potential $\mu_e$ from the charge-equilibrium constraint numerically. We do so by assuming that the pressure of the electrons decouples from that of the quarks and that the electrons can be treated as free particles,
\begin{equation}
    p_e= \frac{1}{12} \left(\frac{\mu_e^4}{\pi^2} +2\mu_e^2 T^2 +\frac{7}{15}\pi^2T^4 \right).
     \label{eq:electronpressure}
\end{equation}
While a simple form of the electron pressure simplifies matters, solving the electron chemical potential remains the most numerically involved part of the calculation. Once it is known, the equilibrium densities and susceptibilities are readily evaluated in both pQCD and the bag model, and their values can be substituted to the formulae for the transport coefficients presented in the main text. 
In addition, there are constraints for the \emph{deviations} $\delta \mu_a$ of chemical 
potentials from their equilibrium values. We impose that these constraints also keep the system
electrically neutral, and that the total baryon number is conserved.
The constraints can be written in terms of the (equilibrium) susceptibilities as
\begin{align}
0&=\delta n_Q =\sum_{a}\left( \chi_e^{\ a} -\chi_u^{\ a}\right) \delta \mu_a, 
\label{eq:charge-constraint}\\
&\,\nonumber\\
0&=\delta n_q =\sum_{a} \left( \chi_d^{\ a} + \chi_s^{\ a}  -  \chi_e^{\ a} \right)\delta \mu_a.
\label{eq:baryon-constraint}
\end{align}
Combined with the definitions of $\mu_{\alpha}$ from \eqref{eq:mu12def} we can obtain the resulting $\delta \mu_{a}$ in terms of the $\mu_{\alpha}$. They, on the other hand, can be substituted into the definition of  bulk scalar \eqref{eq:bulkscalardefn}, which then uniquely gives us the coefficients $\Pi_1,\Pi_2$ used for defining the transport coefficients $\zeta,\xi$. 

As evidenced by the main text, the two equations of state provide us with a unified qualitative picture of the behavior of the transport coefficients.

A standard way of gauging the uncertainty of the pQCD description is by varying the renormalization scale  by a factor of two around a central value, which results in the band shown in the main text (see e.g. \cite{Gorda:2021gha} for details). We note that this also affects the rates, as the coupling and the masses run with the scale (see \cite{Gorda:2021gha} for details). The $\zeta_-$ in Figure \ref{fig:zetaplot}, strongly dependent on the semileptonic rates, shows a considerable uncertainty band, but the other quantities are kept well under check at these densities\footnote{We have also checked that the expected behavior persists with increasing densities, with the band shrinking and the transport quantities gradually shifting.}. This is to be expected from other pQCD computations.

On the other hand, the bag model depends on certain parameters: The bag model pressure has the general form \cite{Torres_2013,Lopes_2021} 
\begin{equation}
    p=p_e+a_4\sum_{f\in \lbrace{u,d,s \rbrace}} p_f + B_\mathrm{eff},
\end{equation}
where $p_e$ is the electron pressure defined above, $p_f$ are the individual quark contributions (each taken to be a free fermion), and $a_4$ and $B_\mathrm{eff}$ are model parameters. Of the quark pressures, the $u,d$ quarks have the same form as the electron pressure, while the massive $s$-quark pressure cannot be expressed in closed form for general $T,\mu_s$, and reads
\begin{widetext}
\begin{equation}  
    p_s=\frac{3 T}{\pi^2} \int_0^\infty  \mathrm{d}k k^2 \, \Bigg\{ \ln\left[1+\exp \left(-\frac{ \sqrt{k^2+m_s^2}-\mu_s}{T} \right)\right]     +\ln\left[1+\exp\left(-\frac{ \sqrt{k^2+m_s^2}+\mu_s}{T} \right) \right] \Bigg\}.
    \label{eq:pressurems}
\end{equation}
\end{widetext}
Here, the $m_s$-parameter should be thought to represent the constituent quark mass, and acts as a third model parameter. Indeed, as the parameter responsible for breaking conformal invariance, it is the one most relevant for bulk viscosities. 
Densities and therefore also susceptibilities are independent of the bag constant, whereas the pressure depends approximately linearly on $a_4$. For small variations of $a_4$, its impact is essentially imperceptible on the doubly logarithmic scale used here for the bulk viscosities.
To be more quantitative, we set $a_4=0.7$ \cite{Ghosh:2025wfx} having checked explicitly that varying it between $0.6-0.8$ changes $\tau_+$ by $\lesssim 15\%$, $\zeta_+$ by $\lesssim 30\%$, and both $\tau_-$ and $\zeta_-$ orders of magnitude less.

\begin{figure}
    \includegraphics[width=0.48\textwidth]{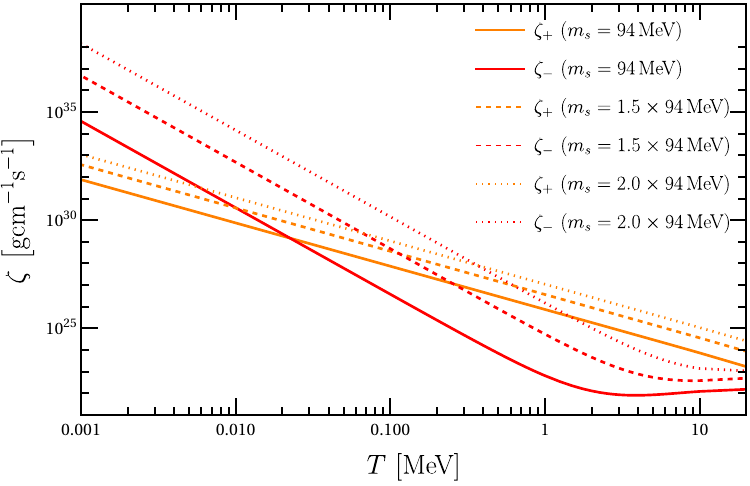}
    \caption{ $\zeta_\pm$ at $m_s\in\lbrace 94,1.5\times 94,2.0\times 94\rbrace\,\mathrm{MeV}$ in the bag model. Everything evaluated at $\mu_d=400\,\mathrm{MeV}, a_4=0.7$.}
    \label{fig:zetamsplot}
\end{figure}

In contrast to $a_4$, varying $m_s$ significantly changes the $\zeta_\pm$ in particular. We show this in Figure \ref{fig:zetamsplot}, displaying the two viscosities for different values of the mass parameter with the $a_4$-parameter fixed to $0.7$. The behavior seen here can be approximated quite well by simple scaling laws, with $\zeta_+$ scaling as $\sim (m_s^2)^2$, and $\zeta_-$ scaling as $\sim (m_s^2)^5$ for small temperatures (below the inflection temperature) and as $\sim (m_s^2)^{3/2}$ above the inflection temperature. Regardless of the sensitivity of the $\zeta$ to $m_s$, we have checked that $T_\mathrm{cross}$ is not significantly affected by the value of $m_s$.

 \section{Explicit formulae}

 \label{sec:explitiformulae}
For the sake of completeness and to possibly give the reader some insight on the explicit form of the bulk scalar
and the transport coefficients, we list them below. We note that as
the forms are rather complicated, it is often simpler to obtain them
using the methods outlined in the main text, as they require merely
simple matrix operations. To simplify formulae, we define the shorthands
\begin{align}
\chiinvdiag{A}{}&\equiv\chiinv{d}{d}-2\chiinv{d}{s}+\chiinv{s}{s} \nonumber, \\
\chiinvdiag{B}{}&\equiv\chiinv{u}{u}-2\chiinv{u}{d}+\chiinv{d}{d}+\chiinv{e}{e} \nonumber, \\
\chiinvdiag{C}{}&\equiv\chiinv{u}{u}-2\chiinv{u}{s}+\chiinv{s}{s}+\chiinv{e}{e}, 
\end{align}
and will make use of the Källén function $K\left(x,y,z\right)=x^{2}+y^{2}+z^{2}-2\left(xy+yz+zx\right)$ \footnote{We opt for the notation $K$ instead of the more common $\lambda$ to avoid confusion with the rates.} and
the pairwise sum $Q\left(x,y,z\right)=xy+yz+zx$. In terms of these,
we find the components of the bulk scalar under the constraints outlined
above to read
\begin{widetext}
\begin{align}
&\Pi_{1}K\left(\chiinvdiag{A}{},\chiinvdiag{B}{},\chiinvdiag{C}{}\right)=n_sK\left(\chiinvdiag{A}{},\chiinvdiag{B}{},\chiinvdiag{C}{}\right)+n_q\nonumber\bigg\lbrace\left[\chiinv{u}{u}+\frac{1}{3}\chiinv{e}{e}\right]\left(\chiinvdiag{A}{}+\chiinvdiag{B}{}-\chiinvdiag{C}{}\right)\nonumber \\
&-\chiinv{d}{d}\left(\chiinvdiag{A}{}-\chiinvdiag{B}{}-\chiinvdiag{C}{}\right)-2\chiinv{s}{s}\chiinvdiag{B}{}+\left(\chiinvdiag{A}{}-\chiinvdiag{B}{}+\chiinvdiag{C}{}\right)\chiinvdiag{B}{}\bigg\rbrace  \\
&\Pi_{2}K\left(\chiinvdiag{A}{},\chiinvdiag{B}{},\chiinvdiag{C}{}\right)=n_uK\left(\chiinvdiag{A}{},\chiinvdiag{B}{},\chiinvdiag{C}{}\right)+n_q\nonumber\bigg\lbrace-2 \left[\chiinv{u}{u}+\frac{1}{3}\chiinv{e}{e}\right]\chiinvdiag{A}{}\nonumber \\
&+\chiinv{d}{d}\left(\chiinvdiag{A}{}-\chiinvdiag{B}{}+\chiinvdiag{C}{}\right)+\chiinv{s}{s}\left(\chiinvdiag{A}{}+\chiinvdiag{B}{}-\chiinvdiag{C}{}\right)-\left(\chiinvdiag{A}{}-\chiinvdiag{B}{}-\chiinvdiag{C}{}\right)\chiinvdiag{A}{}\bigg\rbrace 
\end{align}
  \end{widetext}
whereas the (inverse) relaxation times are
\begin{widetext}
\begin{align}
2\tau_{\pm}^{-1}=\left(\chiinvdiag{A}{}\lambda_{1}+\chiinvdiag{B}{}\lambda_{2}+\chiinvdiag{C}{}\lambda_{3}\right)&\pm\sqrt{\left(\chiinvdiag{A}{}\lambda_{1}+\chiinvdiag{B}{}\lambda_{2}+\chiinvdiag{C}{}\lambda_{3}\right)^{2}+Q\left(\lambda_{1},\lambda_{2},\lambda_{3}\right)K\left(\chiinvdiag{A}{},\chiinvdiag{B}{},\chiinvdiag{C}{}\right)}.
\end{align}
  \end{widetext}
Lastly,  the coefficients $\zeta,\xi$ are given by 
  \begin{widetext}
\begin{align}
&\zeta Q\left(\lambda_{1},\lambda_{2},\lambda_{3}\right)K\left(\chi_{A}^{-1},\chi_{B}^{-1},\chi_{C}^{-1}\right)=\left[\left(n_{u}+n_{s}\right)\lambda_{3}\left(\Pi_{1}+\Pi_{2}\right)+n_{s}\lambda_{2}\Pi_{1}+n_{u}\lambda_{1}\Pi_{2}\right]K\left(\chi_{A}^{-1},\chi_{B}^{-1},\chi_{C}^{-1}\right)\nonumber \\
&+\Bigg\{ -\left(\chi_{A}^{-1}-\chi_{B}^{-1}\right)^{2}+\left(\chi_{A}^{-1}+\chi_{B}^{-1}\right)\chi_{C}^{-1}-\left(\chi_{A}^{-1}-\chi_{B}^{-1}+\chi_{C}^{-1}\right)\left[\left(\chi_{u}^{\ u}\right)^{-1}+\frac{1}{3}\left(\chi_{e}^{\ e}\right)^{-1}\right]\nonumber \\
&+2\chi_{C}^{-1}\left(\chi_{d}^{\ d}\right)^{-1}+\left(\chi_{A}^{-1}-\chi_{B}^{-1}-\chi_{C}^{-1}\right)\left(\chi_{s}^{\ s}\right)^{-1}\Bigg\} n_q\lambda_{3}\left(\Pi_{1}+\Pi_{2}\right)\nonumber \\
&+\Bigg\{ \left(\chi_{A}^{-1}-\chi_{B}^{-1}+\chi_{C}^{-1}\right)\chi_{B}^{-1}+\left(\chi_{A}^{-1}+\chi_{B}^{-1}-\chi_{C}^{-1}\right)\left[\left(\chi_{u}^{\ u}\right)^{-1}+\frac{1}{3}\left(\chi_{e}^{\ e}\right)^{-1}\right]\nonumber \\
&-\left(\chi_{A}^{-1}-\chi_{B}^{-1}-\chi_{C}^{-1}\right)\left(\chi_{d}^{\ d}\right)^{-1}-2\chi_{B}^{-1}\left(\chi_{s}^{\ s}\right)^{-1}\Bigg\} n_q\lambda_{2}\Pi_{1}\nonumber \\
&+\Bigg\{- \left(\chi_{A}^{-1}-\chi_{B}^{-1}-\chi_{C}^{-1}\right)\chi_{A}^{-1}-2\chi_{A}^{-1}\left[\left(\chi_{u}^{\ u}\right)^{-1}+\frac{1}{3}\left(\chi_{e}^{\ e}\right)^{-1}\right]\nonumber \\
&+\left(\chi_{A}^{-1}-\chi_{B}^{-1}+\chi_{C}^{-1}\right)\left(\chi_{d}^{\ d}\right)^{-1}+\left(\chi_{A}^{-1}+\chi_{B}^{-1}-\chi_{C}^{-1}\right)\left(\chi_{s}^{\ s}\right)^{-1}\Bigg\} n_q\lambda_{1}\Pi_{2}, \\
&\frac{\xi}{2}Q\left(\lambda_{1},\lambda_{2},\lambda_{3}\right)K\left(\chiinvdiag{A}{},\chiinvdiag{B}{},\chiinvdiag{C}{}\right)=\left\{ n_q\left[\chiinv{d}{d}-\chiinv{s}{s}\right]+\left(n_d-n_s\right)\chiinvdiag{A}{}-n_{u}\left(\chiinvdiag{B}{}-\chiinvdiag{C}{}\right)\right\} \Pi_{1} \nonumber \\
&+\left\{ n_q\left[\chiinv{d}{d}-\chiinv{u}{u}-\frac{1}{3}\chiinv{e}{e}\right]+\left(n_d-n_u\right)\chiinvdiag{B}{}-n_{s}\left(\chiinvdiag{A}{}-\chiinvdiag{C}{}\right)\right\} \Pi_{2},
\end{align}
  \end{widetext}
from which one may evaluate $\zeta_{\pm}$ using 
Eq.~(\ref{eq:zetapm}).

\bibliography{bibliography}

\begin{thebibliography}{60}%
\makeatletter
\providecommand \@ifxundefined [1]{%
 \@ifx{#1\undefined}
}%
\providecommand \@ifnum [1]{%
 \ifnum #1\expandafter \@firstoftwo
 \else \expandafter \@secondoftwo
 \fi
}%
\providecommand \@ifx [1]{%
 \ifx #1\expandafter \@firstoftwo
 \else \expandafter \@secondoftwo
 \fi
}%
\providecommand \natexlab [1]{#1}%
\providecommand \enquote  [1]{``#1''}%
\providecommand \bibnamefont  [1]{#1}%
\providecommand \bibfnamefont [1]{#1}%
\providecommand \citenamefont [1]{#1}%
\providecommand \href@noop [0]{\@secondoftwo}%
\providecommand \href [0]{\begingroup \@sanitize@url \@href}%
\providecommand \@href[1]{\@@startlink{#1}\@@href}%
\providecommand \@@href[1]{\endgroup#1\@@endlink}%
\providecommand \@sanitize@url [0]{\catcode `\\12\catcode `\$12\catcode
  `\&12\catcode `\#12\catcode `\^12\catcode `\_12\catcode `\%12\relax}%
\providecommand \@@startlink[1]{}%
\providecommand \@@endlink[0]{}%
\providecommand \url  [0]{\begingroup\@sanitize@url \@url }%
\providecommand \@url [1]{\endgroup\@href {#1}{\urlprefix }}%
\providecommand \urlprefix  [0]{URL }%
\providecommand \Eprint [0]{\href }%
\providecommand \doibase [0]{https://doi.org/}%
\providecommand \selectlanguage [0]{\@gobble}%
\providecommand \bibinfo  [0]{\@secondoftwo}%
\providecommand \bibfield  [0]{\@secondoftwo}%
\providecommand \translation [1]{[#1]}%
\providecommand \BibitemOpen [0]{}%
\providecommand \bibitemStop [0]{}%
\providecommand \bibitemNoStop [0]{.\EOS\space}%
\providecommand \EOS [0]{\spacefactor3000\relax}%
\providecommand \BibitemShut  [1]{\csname bibitem#1\endcsname}%
\let\auto@bib@innerbib\@empty
\bibitem [{\citenamefont {Kogut}\ and\ \citenamefont
  {Stephanov}(2009)}]{qcdbook}%
  \BibitemOpen
  \bibfield  {author} {\bibinfo {author} {\bibfnamefont {J.}~\bibnamefont
  {Kogut}}\ and\ \bibinfo {author} {\bibfnamefont {M.}~\bibnamefont
  {Stephanov}},\ }\href@noop {} {\emph {\bibinfo {title} {The Phases of Quantum
  Chromodynamics}}}\ (\bibinfo  {publisher} {Cambridge University Press},\
  \bibinfo {year} {2009})\BibitemShut {NoStop}%
\bibitem [{\citenamefont {Weber}(2005)}]{WEBER2005193}%
  \BibitemOpen
  \bibfield  {author} {\bibinfo {author} {\bibfnamefont {F.}~\bibnamefont
  {Weber}},\ }\bibfield  {title} {\bibinfo {title} {Strange quark matter and
  compact stars},\ }\href {https://doi.org/10.1016/j.ppnp.2004.07.001}
  {\bibfield  {journal} {\bibinfo  {journal} {Prog. Part. Nucl. Phys.}\
  }\textbf {\bibinfo {volume} {54}},\ \bibinfo {pages} {193} (\bibinfo {year}
  {2005})},\ \Eprint {https://arxiv.org/abs/astro-ph/0407155}
  {arXiv:astro-ph/0407155} \BibitemShut {NoStop}%
\bibitem [{\citenamefont {Alford}\ \emph {et~al.}(2008)\citenamefont {Alford},
  \citenamefont {Schmitt}, \citenamefont {Rajagopal},\ and\ \citenamefont
  {Sch{\"a}fer}}]{Alford_2008}%
  \BibitemOpen
  \bibfield  {author} {\bibinfo {author} {\bibfnamefont {M.~G.}\ \bibnamefont
  {Alford}}, \bibinfo {author} {\bibfnamefont {A.}~\bibnamefont {Schmitt}},
  \bibinfo {author} {\bibfnamefont {K.}~\bibnamefont {Rajagopal}},\ and\
  \bibinfo {author} {\bibfnamefont {T.}~\bibnamefont {Sch{\"a}fer}},\
  }\bibfield  {title} {\bibinfo {title} {Color superconductivity in dense quark
  matter},\ }\href {https://doi.org/10.1103/RevModPhys.80.1455} {\bibfield
  {journal} {\bibinfo  {journal} {Rev. Mod. Phys.}\ }\textbf {\bibinfo {volume}
  {80}},\ \bibinfo {pages} {1455} (\bibinfo {year} {2008})},\ \Eprint
  {https://arxiv.org/abs/0709.4635} {arXiv:0709.4635 [hep-ph]} \BibitemShut
  {NoStop}%
\bibitem [{\citenamefont {Alford}\ \emph {et~al.}(2019)\citenamefont {Alford},
  \citenamefont {Han},\ and\ \citenamefont {Schwenzer}}]{Alford:2019oge}%
  \BibitemOpen
  \bibfield  {author} {\bibinfo {author} {\bibfnamefont {M.~G.}\ \bibnamefont
  {Alford}}, \bibinfo {author} {\bibfnamefont {S.}~\bibnamefont {Han}},\ and\
  \bibinfo {author} {\bibfnamefont {K.}~\bibnamefont {Schwenzer}},\ }\bibfield
  {title} {\bibinfo {title} {Signatures for quark matter from multi-messenger
  observations},\ }\href {https://doi.org/10.1088/1361-6471/ab337a} {\bibfield
  {journal} {\bibinfo  {journal} {J. Phys. G}\ }\textbf {\bibinfo {volume}
  {46}},\ \bibinfo {pages} {114001} (\bibinfo {year} {2019})},\ \Eprint
  {https://arxiv.org/abs/1904.05471} {arXiv:1904.05471 [nucl-th]} \BibitemShut
  {NoStop}%
\bibitem [{\citenamefont {Abbott}\ \emph
  {et~al.}(2017{\natexlab{a}})\citenamefont {Abbott} \emph
  {et~al.}}]{LIGOScientific:2017vwq}%
  \BibitemOpen
  \bibfield  {author} {\bibinfo {author} {\bibfnamefont {B.~P.}\ \bibnamefont
  {Abbott}} \emph {et~al.} (\bibinfo {collaboration} {LIGO Scientific,
  Virgo}),\ }\bibfield  {title} {\bibinfo {title} {Gw170817: Observation of
  gravitational waves from a binary neutron star inspiral},\ }\href
  {https://doi.org/10.1103/PhysRevLett.119.161101} {\bibfield  {journal}
  {\bibinfo  {journal} {Phys. Rev. Lett.}\ }\textbf {\bibinfo {volume} {119}},\
  \bibinfo {pages} {161101} (\bibinfo {year} {2017}{\natexlab{a}})},\ \Eprint
  {https://arxiv.org/abs/1710.05832} {arXiv:1710.05832 [gr-qc]} \BibitemShut
  {NoStop}%
\bibitem [{\citenamefont {Abbott}\ \emph
  {et~al.}(2017{\natexlab{b}})\citenamefont {Abbott} \emph
  {et~al.}}]{LIGOScientific:2017ync}%
  \BibitemOpen
  \bibfield  {author} {\bibinfo {author} {\bibfnamefont {B.~P.}\ \bibnamefont
  {Abbott}} \emph {et~al.} (\bibinfo {collaboration} {LIGO Scientific, Virgo,
  Fermi GBM, INTEGRAL, IceCube, AstroSat Cadmium Zinc Telluride Imager Team,
  IPN, Insight-Hxmt, ANTARES, Swift, AGILE Team, 1M2H Team, Dark Energy Camera
  GW-EM, DES, DLT40, GRAWITA, Fermi-LAT, ATCA, ASKAP, Las Cumbres Observatory
  Group, OzGrav, DWF (Deeper Wider Faster Program), AST3, CAASTRO, VINROUGE,
  MASTER, J-GEM, GROWTH, JAGWAR, CaltechNRAO, TTU-NRAO, NuSTAR, Pan-STARRS,
  MAXI Team, TZAC Consortium, KU, Nordic Optical Telescope, ePESSTO, GROND,
  Texas Tech University, SALT Group, TOROS, BOOTES, MWA, CALET, IKI-GW
  Follow-up, H.E.S.S., LOFAR, LWA, HAWC, Pierre Auger, ALMA, Euro VLBI Team, Pi
  of Sky, Chandra Team at McGill University, DFN, ATLAS Telescopes, High Time
  Resolution Universe Survey, RIMAS, RATIR, SKA South Africa/MeerKAT}),\
  }\bibfield  {title} {\bibinfo {title} {Multi-messenger observations of a
  binary neutron star merger},\ }\href
  {https://doi.org/10.3847/2041-8213/aa91c9} {\bibfield  {journal} {\bibinfo
  {journal} {Astrophys. J. Lett.}\ }\textbf {\bibinfo {volume} {848}},\
  \bibinfo {pages} {L12} (\bibinfo {year} {2017}{\natexlab{b}})},\ \Eprint
  {https://arxiv.org/abs/1710.05833} {arXiv:1710.05833 [astro-ph.HE]}
  \BibitemShut {NoStop}%
\bibitem [{\citenamefont {Faber}\ and\ \citenamefont
  {Rasio}(2012)}]{Faber:2012rw}%
  \BibitemOpen
  \bibfield  {author} {\bibinfo {author} {\bibfnamefont {J.~A.}\ \bibnamefont
  {Faber}}\ and\ \bibinfo {author} {\bibfnamefont {F.~A.}\ \bibnamefont
  {Rasio}},\ }\bibfield  {title} {\bibinfo {title} {Binary neutron star
  mergers},\ }\href {https://doi.org/10.12942/lrr-2012-8} {\bibfield  {journal}
  {\bibinfo  {journal} {Living Rev. Rel.}\ }\textbf {\bibinfo {volume} {15}},\
  \bibinfo {pages} {8} (\bibinfo {year} {2012})},\ \Eprint
  {https://arxiv.org/abs/1204.3858} {arXiv:1204.3858 [gr-qc]} \BibitemShut
  {NoStop}%
\bibitem [{\citenamefont {Baiotti}\ and\ \citenamefont
  {Rezzolla}(2017)}]{Baiotti:2016qnr}%
  \BibitemOpen
  \bibfield  {author} {\bibinfo {author} {\bibfnamefont {L.}~\bibnamefont
  {Baiotti}}\ and\ \bibinfo {author} {\bibfnamefont {L.}~\bibnamefont
  {Rezzolla}},\ }\bibfield  {title} {\bibinfo {title} {Binary neutron star
  mergers: a review of einstein\textquoteright{}s richest laboratory},\ }\href
  {https://doi.org/10.1088/1361-6633/aa67bb} {\bibfield  {journal} {\bibinfo
  {journal} {Rept. Prog. Phys.}\ }\textbf {\bibinfo {volume} {80}},\ \bibinfo
  {pages} {096901} (\bibinfo {year} {2017})},\ \Eprint
  {https://arxiv.org/abs/1607.03540} {arXiv:1607.03540 [gr-qc]} \BibitemShut
  {NoStop}%
\bibitem [{\citenamefont {Most}\ \emph {et~al.}(2024)\citenamefont {Most},
  \citenamefont {Haber}, \citenamefont {Harris}, \citenamefont {Zhang},
  \citenamefont {Alford},\ and\ \citenamefont {Noronha}}]{Most:2022yhe}%
  \BibitemOpen
  \bibfield  {author} {\bibinfo {author} {\bibfnamefont {E.~R.}\ \bibnamefont
  {Most}}, \bibinfo {author} {\bibfnamefont {A.}~\bibnamefont {Haber}},
  \bibinfo {author} {\bibfnamefont {S.~P.}\ \bibnamefont {Harris}}, \bibinfo
  {author} {\bibfnamefont {Z.}~\bibnamefont {Zhang}}, \bibinfo {author}
  {\bibfnamefont {M.~G.}\ \bibnamefont {Alford}},\ and\ \bibinfo {author}
  {\bibfnamefont {J.}~\bibnamefont {Noronha}},\ }\bibfield  {title} {\bibinfo
  {title} {Emergence of microphysical bulk viscosity in binary neutron star
  postmerger dynamics},\ }\href {https://doi.org/10.3847/2041-8213/ad454f}
  {\bibfield  {journal} {\bibinfo  {journal} {Astrophys. J. Lett.}\ }\textbf
  {\bibinfo {volume} {967}},\ \bibinfo {pages} {L14} (\bibinfo {year}
  {2024})},\ \Eprint {https://arxiv.org/abs/2207.00442} {arXiv:2207.00442
  [astro-ph.HE]} \BibitemShut {NoStop}%
\bibitem [{\citenamefont {Chabanov}\ and\ \citenamefont
  {Rezzolla}(2025{\natexlab{a}})}]{Chabanov:2023blf}%
  \BibitemOpen
  \bibfield  {author} {\bibinfo {author} {\bibfnamefont {M.}~\bibnamefont
  {Chabanov}}\ and\ \bibinfo {author} {\bibfnamefont {L.}~\bibnamefont
  {Rezzolla}},\ }\bibfield  {title} {\bibinfo {title} {Impact of bulk viscosity
  on the postmerger gravitational-wave signal from merging neutron stars},\
  }\href {https://doi.org/10.1103/PhysRevLett.134.071402} {\bibfield  {journal}
  {\bibinfo  {journal} {Phys. Rev. Lett.}\ }\textbf {\bibinfo {volume} {134}},\
  \bibinfo {pages} {071402} (\bibinfo {year} {2025}{\natexlab{a}})},\ \Eprint
  {https://arxiv.org/abs/2307.10464} {arXiv:2307.10464 [gr-qc]} \BibitemShut
  {NoStop}%
\bibitem [{\citenamefont {Chabanov}\ and\ \citenamefont
  {Rezzolla}(2025{\natexlab{b}})}]{Chabanov:2023abq}%
  \BibitemOpen
  \bibfield  {author} {\bibinfo {author} {\bibfnamefont {M.}~\bibnamefont
  {Chabanov}}\ and\ \bibinfo {author} {\bibfnamefont {L.}~\bibnamefont
  {Rezzolla}},\ }\bibfield  {title} {\bibinfo {title} {Numerical modeling of
  bulk viscosity in neutron stars},\ }\href
  {https://doi.org/10.1103/PhysRevD.111.044074} {\bibfield  {journal} {\bibinfo
   {journal} {Phys. Rev. D}\ }\textbf {\bibinfo {volume} {111}},\ \bibinfo
  {pages} {044074} (\bibinfo {year} {2025}{\natexlab{b}})},\ \Eprint
  {https://arxiv.org/abs/2311.13027} {arXiv:2311.13027 [gr-qc]} \BibitemShut
  {NoStop}%
\bibitem [{\citenamefont {Kokkotas}\ and\ \citenamefont
  {Schmidt}(1999)}]{Kokkotas:1999bd}%
  \BibitemOpen
  \bibfield  {author} {\bibinfo {author} {\bibfnamefont {K.~D.}\ \bibnamefont
  {Kokkotas}}\ and\ \bibinfo {author} {\bibfnamefont {B.~G.}\ \bibnamefont
  {Schmidt}},\ }\bibfield  {title} {\bibinfo {title} {Quasinormal modes of
  stars and black holes},\ }\href {https://doi.org/10.12942/lrr-1999-2}
  {\bibfield  {journal} {\bibinfo  {journal} {Living Rev. Rel.}\ }\textbf
  {\bibinfo {volume} {2}},\ \bibinfo {pages} {2} (\bibinfo {year} {1999})},\
  \Eprint {https://arxiv.org/abs/gr-qc/9909058} {arXiv:gr-qc/9909058}
  \BibitemShut {NoStop}%
\bibitem [{\citenamefont {Rezzolla}(2003)}]{Rezzolla:2003ua}%
  \BibitemOpen
  \bibfield  {author} {\bibinfo {author} {\bibfnamefont {L.}~\bibnamefont
  {Rezzolla}},\ }\bibfield  {title} {\bibinfo {title} {Gravitational waves from
  perturbed black holes and relativistic stars},\ }\href@noop {} {\bibfield
  {journal} {\bibinfo  {journal} {ICTP Lect. Notes Ser.}\ }\textbf {\bibinfo
  {volume} {14}},\ \bibinfo {pages} {255} (\bibinfo {year} {2003})},\ \Eprint
  {https://arxiv.org/abs/gr-qc/0302025} {arXiv:gr-qc/0302025} \BibitemShut
  {NoStop}%
\bibitem [{\citenamefont {Ghosh}\ \emph {et~al.}(2025)\citenamefont {Ghosh},
  \citenamefont {Hern\'andez}, \citenamefont {Pradhan}, \citenamefont {Manuel},
  \citenamefont {Chatterjee},\ and\ \citenamefont {Tolos}}]{Ghosh:2025wfx}%
  \BibitemOpen
  \bibfield  {author} {\bibinfo {author} {\bibfnamefont {S.}~\bibnamefont
  {Ghosh}}, \bibinfo {author} {\bibfnamefont {J.~L.}\ \bibnamefont
  {Hern\'andez}}, \bibinfo {author} {\bibfnamefont {B.~K.}\ \bibnamefont
  {Pradhan}}, \bibinfo {author} {\bibfnamefont {C.}~\bibnamefont {Manuel}},
  \bibinfo {author} {\bibfnamefont {D.}~\bibnamefont {Chatterjee}},\ and\
  \bibinfo {author} {\bibfnamefont {L.}~\bibnamefont {Tolos}},\ }\href@noop {}
  {\bibinfo {title} {Tidal heating in binary inspiral of strange quark stars}}
  (\bibinfo {year} {2025}),\ \Eprint {https://arxiv.org/abs/2504.07659}
  {arXiv:2504.07659 [gr-qc]} \BibitemShut {NoStop}%
\bibitem [{\citenamefont {Sawyer}(1989{\natexlab{a}})}]{Sawyer:1989dp}%
  \BibitemOpen
  \bibfield  {author} {\bibinfo {author} {\bibfnamefont {R.~F.}\ \bibnamefont
  {Sawyer}},\ }\bibfield  {title} {\bibinfo {title} {Bulk viscosity of hot
  neutron-star matter and the maximum rotation rates of neutron stars},\ }\href
  {https://doi.org/10.1103/PhysRevD.39.3804} {\bibfield  {journal} {\bibinfo
  {journal} {Phys. Rev. D}\ }\textbf {\bibinfo {volume} {39}},\ \bibinfo
  {pages} {3804} (\bibinfo {year} {1989}{\natexlab{a}})}\BibitemShut {NoStop}%
\bibitem [{\citenamefont {Haensel}\ and\ \citenamefont
  {Schaeffer}(1992)}]{Haensel:1992zz}%
  \BibitemOpen
  \bibfield  {author} {\bibinfo {author} {\bibfnamefont {P.}~\bibnamefont
  {Haensel}}\ and\ \bibinfo {author} {\bibfnamefont {R.}~\bibnamefont
  {Schaeffer}},\ }\bibfield  {title} {\bibinfo {title} {Bulk viscosity of
  hot-neutron-star matter from direct urca processes},\ }\href
  {https://doi.org/10.1103/PhysRevD.45.4708} {\bibfield  {journal} {\bibinfo
  {journal} {Phys. Rev. D}\ }\textbf {\bibinfo {volume} {45}},\ \bibinfo
  {pages} {4708} (\bibinfo {year} {1992})}\BibitemShut {NoStop}%
\bibitem [{\citenamefont {Haensel}\ \emph {et~al.}(2000)\citenamefont
  {Haensel}, \citenamefont {Levenfish},\ and\ \citenamefont
  {Yakovlev}}]{Haensel:2000vz}%
  \BibitemOpen
  \bibfield  {author} {\bibinfo {author} {\bibfnamefont {P.}~\bibnamefont
  {Haensel}}, \bibinfo {author} {\bibfnamefont {K.~P.}\ \bibnamefont
  {Levenfish}},\ and\ \bibinfo {author} {\bibfnamefont {D.~G.}\ \bibnamefont
  {Yakovlev}},\ }\bibfield  {title} {\bibinfo {title} {Bulk viscosity in
  superfluid neutron star cores. i. direct urca processes in npe mu matter},\
  }\href@noop {} {\bibfield  {journal} {\bibinfo  {journal} {Astron.
  Astrophys.}\ }\textbf {\bibinfo {volume} {357}},\ \bibinfo {pages} {1157}
  (\bibinfo {year} {2000})},\ \Eprint {https://arxiv.org/abs/astro-ph/0004183}
  {arXiv:astro-ph/0004183} \BibitemShut {NoStop}%
\bibitem [{\citenamefont {Alford}\ and\ \citenamefont
  {Harris}(2018)}]{Alford:2018lhf}%
  \BibitemOpen
  \bibfield  {author} {\bibinfo {author} {\bibfnamefont {M.~G.}\ \bibnamefont
  {Alford}}\ and\ \bibinfo {author} {\bibfnamefont {S.~P.}\ \bibnamefont
  {Harris}},\ }\bibfield  {title} {\bibinfo {title} {Beta equilibrium in
  neutron star mergers},\ }\href {https://doi.org/10.1103/PhysRevC.98.065806}
  {\bibfield  {journal} {\bibinfo  {journal} {Phys. Rev. C}\ }\textbf {\bibinfo
  {volume} {98}},\ \bibinfo {pages} {065806} (\bibinfo {year} {2018})},\
  \Eprint {https://arxiv.org/abs/1803.00662} {arXiv:1803.00662 [nucl-th]}
  \BibitemShut {NoStop}%
\bibitem [{\citenamefont {Alford}\ and\ \citenamefont
  {Harris}(2019)}]{Alford:2019qtm}%
  \BibitemOpen
  \bibfield  {author} {\bibinfo {author} {\bibfnamefont {M.~G.}\ \bibnamefont
  {Alford}}\ and\ \bibinfo {author} {\bibfnamefont {S.~P.}\ \bibnamefont
  {Harris}},\ }\bibfield  {title} {\bibinfo {title} {Damping of density
  oscillations in neutrino-transparent nuclear matter},\ }\href
  {https://doi.org/10.1103/PhysRevC.100.035803} {\bibfield  {journal} {\bibinfo
   {journal} {Phys. Rev. C}\ }\textbf {\bibinfo {volume} {100}},\ \bibinfo
  {pages} {035803} (\bibinfo {year} {2019})},\ \Eprint
  {https://arxiv.org/abs/1907.03795} {arXiv:1907.03795 [nucl-th]} \BibitemShut
  {NoStop}%
\bibitem [{\citenamefont {Alford}\ \emph
  {et~al.}(2024{\natexlab{a}})\citenamefont {Alford}, \citenamefont {Haber},\
  and\ \citenamefont {Zhang}}]{Alford:2023gxq}%
  \BibitemOpen
  \bibfield  {author} {\bibinfo {author} {\bibfnamefont {M.~G.}\ \bibnamefont
  {Alford}}, \bibinfo {author} {\bibfnamefont {A.}~\bibnamefont {Haber}},\ and\
  \bibinfo {author} {\bibfnamefont {Z.}~\bibnamefont {Zhang}},\ }\bibfield
  {title} {\bibinfo {title} {Isospin equilibration in neutron star mergers},\
  }\href {https://doi.org/10.1103/PhysRevC.109.055803} {\bibfield  {journal}
  {\bibinfo  {journal} {Phys. Rev. C}\ }\textbf {\bibinfo {volume} {109}},\
  \bibinfo {pages} {055803} (\bibinfo {year} {2024}{\natexlab{a}})},\ \Eprint
  {https://arxiv.org/abs/2306.06180} {arXiv:2306.06180 [nucl-th]} \BibitemShut
  {NoStop}%
\bibitem [{\citenamefont {Sawyer}(1989{\natexlab{b}})}]{Sawyer:1989uy}%
  \BibitemOpen
  \bibfield  {author} {\bibinfo {author} {\bibfnamefont {R.~F.}\ \bibnamefont
  {Sawyer}},\ }\bibfield  {title} {\bibinfo {title} {Damping of vibrations and
  of the secular instability in quark stars},\ }\href
  {https://doi.org/10.1016/0370-2693(95)00145-B} {\bibfield  {journal}
  {\bibinfo  {journal} {Phys. Lett. B}\ }\textbf {\bibinfo {volume} {233}},\
  \bibinfo {pages} {412} (\bibinfo {year} {1989}{\natexlab{b}})},\ \bibinfo
  {note} {[Erratum: Phys.Lett.B 237, 605 (1990), Erratum: Phys.Lett.B 347,
  467--467 (1995)]}\BibitemShut {NoStop}%
\bibitem [{\citenamefont {Madsen}(1992)}]{Madsen:1992sx}%
  \BibitemOpen
  \bibfield  {author} {\bibinfo {author} {\bibfnamefont {J.}~\bibnamefont
  {Madsen}},\ }\bibfield  {title} {\bibinfo {title} {Bulk viscosity of strange
  quark matter, damping of quark star vibration, and the maximum rotation rate
  of pulsars},\ }\href {https://doi.org/10.1103/PhysRevD.46.3290} {\bibfield
  {journal} {\bibinfo  {journal} {Phys. Rev. D}\ }\textbf {\bibinfo {volume}
  {46}},\ \bibinfo {pages} {3290} (\bibinfo {year} {1992})}\BibitemShut
  {NoStop}%
\bibitem [{\citenamefont {Sa'd}\ \emph {et~al.}(2007)\citenamefont {Sa'd},
  \citenamefont {Shovkovy},\ and\ \citenamefont {Rischke}}]{Sad:2007afd}%
  \BibitemOpen
  \bibfield  {author} {\bibinfo {author} {\bibfnamefont {B.~A.}\ \bibnamefont
  {Sa'd}}, \bibinfo {author} {\bibfnamefont {I.~A.}\ \bibnamefont {Shovkovy}},\
  and\ \bibinfo {author} {\bibfnamefont {D.~H.}\ \bibnamefont {Rischke}},\
  }\bibfield  {title} {\bibinfo {title} {Bulk viscosity of strange quark
  matter: Urca versus nonleptonic processes},\ }\href
  {https://doi.org/10.1103/PhysRevD.75.125004} {\bibfield  {journal} {\bibinfo
  {journal} {Phys. Rev. D}\ }\textbf {\bibinfo {volume} {75}},\ \bibinfo
  {pages} {125004} (\bibinfo {year} {2007})}\BibitemShut {NoStop}%
\bibitem [{\citenamefont {Cruz~Rojas}\ \emph {et~al.}(2024)\citenamefont
  {Cruz~Rojas}, \citenamefont {Gorda}, \citenamefont {Hoyos}, \citenamefont
  {Jokela}, \citenamefont {J\"arvinen}, \citenamefont {Kurkela}, \citenamefont
  {Paatelainen}, \citenamefont {S\"appi},\ and\ \citenamefont
  {Vuorinen}}]{CruzRojas:2024etx}%
  \BibitemOpen
  \bibfield  {author} {\bibinfo {author} {\bibfnamefont {J.}~\bibnamefont
  {Cruz~Rojas}}, \bibinfo {author} {\bibfnamefont {T.}~\bibnamefont {Gorda}},
  \bibinfo {author} {\bibfnamefont {C.}~\bibnamefont {Hoyos}}, \bibinfo
  {author} {\bibfnamefont {N.}~\bibnamefont {Jokela}}, \bibinfo {author}
  {\bibfnamefont {M.}~\bibnamefont {J\"arvinen}}, \bibinfo {author}
  {\bibfnamefont {A.}~\bibnamefont {Kurkela}}, \bibinfo {author} {\bibfnamefont
  {R.}~\bibnamefont {Paatelainen}}, \bibinfo {author} {\bibfnamefont
  {S.}~\bibnamefont {S\"appi}},\ and\ \bibinfo {author} {\bibfnamefont
  {A.}~\bibnamefont {Vuorinen}},\ }\bibfield  {title} {\bibinfo {title}
  {Estimate for the bulk viscosity of strongly coupled quark matter using
  perturbative qcd and holography},\ }\href
  {https://doi.org/10.1103/PhysRevLett.133.071901} {\bibfield  {journal}
  {\bibinfo  {journal} {Phys. Rev. Lett.}\ }\textbf {\bibinfo {volume} {133}},\
  \bibinfo {pages} {071901} (\bibinfo {year} {2024})},\ \Eprint
  {https://arxiv.org/abs/2402.00621} {arXiv:2402.00621 [hep-ph]} \BibitemShut
  {NoStop}%
\bibitem [{\citenamefont {Hernandez}\ \emph {et~al.}(2024)\citenamefont
  {Hernandez}, \citenamefont {Manuel},\ and\ \citenamefont
  {Tolos}}]{Hernandez:2024rxi}%
  \BibitemOpen
  \bibfield  {author} {\bibinfo {author} {\bibfnamefont {J.~L.}\ \bibnamefont
  {Hernandez}}, \bibinfo {author} {\bibfnamefont {C.}~\bibnamefont {Manuel}},\
  and\ \bibinfo {author} {\bibfnamefont {L.}~\bibnamefont {Tolos}},\ }\bibfield
   {title} {\bibinfo {title} {Damping of density oscillations from bulk
  viscosity in quark matter},\ }\href
  {https://doi.org/10.1103/PhysRevD.109.123022} {\bibfield  {journal} {\bibinfo
   {journal} {Phys. Rev. D}\ }\textbf {\bibinfo {volume} {109}},\ \bibinfo
  {pages} {123022} (\bibinfo {year} {2024})},\ \Eprint
  {https://arxiv.org/abs/2402.06595} {arXiv:2402.06595 [hep-ph]} \BibitemShut
  {NoStop}%
\bibitem [{\citenamefont {Harris}(2024)}]{Harris:2024evy}%
  \BibitemOpen
  \bibfield  {author} {\bibinfo {author} {\bibfnamefont {S.~P.}\ \bibnamefont
  {Harris}},\ }\bibinfo {title} {Nuclear theory in the age of multimessenger
  astronomy}\ (\bibinfo  {publisher} {CRC Press},\ \bibinfo {year} {2024})\
  Chap.\ \bibinfo {chapter} {Bulk Viscosity in Dense Nuclear Matter},\ \Eprint
  {https://arxiv.org/abs/2407.16157} {arXiv:2407.16157 [nucl-th]} \BibitemShut
  {NoStop}%
\bibitem [{\citenamefont {Israel}\ and\ \citenamefont
  {Stewart}(1979)}]{Israel:1979wp}%
  \BibitemOpen
  \bibfield  {author} {\bibinfo {author} {\bibfnamefont {W.}~\bibnamefont
  {Israel}}\ and\ \bibinfo {author} {\bibfnamefont {J.~M.}\ \bibnamefont
  {Stewart}},\ }\bibfield  {title} {\bibinfo {title} {Transient relativistic
  thermodynamics and kinetic theory},\ }\href
  {https://doi.org/10.1016/0003-4916(79)90130-1} {\bibfield  {journal}
  {\bibinfo  {journal} {Annals Phys.}\ }\textbf {\bibinfo {volume} {118}},\
  \bibinfo {pages} {341} (\bibinfo {year} {1979})}\BibitemShut {NoStop}%
\bibitem [{\citenamefont {Hiscock}\ and\ \citenamefont
  {Lindblom}(1983)}]{Hiscock:1983zz}%
  \BibitemOpen
  \bibfield  {author} {\bibinfo {author} {\bibfnamefont {W.~A.}\ \bibnamefont
  {Hiscock}}\ and\ \bibinfo {author} {\bibfnamefont {L.}~\bibnamefont
  {Lindblom}},\ }\bibfield  {title} {\bibinfo {title} {Stability and causality
  in dissipative relativistic fluids},\ }\href
  {https://doi.org/10.1016/0003-4916(83)90288-9} {\bibfield  {journal}
  {\bibinfo  {journal} {Annals Phys.}\ }\textbf {\bibinfo {volume} {151}},\
  \bibinfo {pages} {466} (\bibinfo {year} {1983})}\BibitemShut {NoStop}%
\bibitem [{\citenamefont {Bemfica}\ \emph {et~al.}(2019)\citenamefont
  {Bemfica}, \citenamefont {Disconzi},\ and\ \citenamefont
  {Noronha}}]{Bemfica:2019cop}%
  \BibitemOpen
  \bibfield  {author} {\bibinfo {author} {\bibfnamefont {F.~S.}\ \bibnamefont
  {Bemfica}}, \bibinfo {author} {\bibfnamefont {M.~M.}\ \bibnamefont
  {Disconzi}},\ and\ \bibinfo {author} {\bibfnamefont {J.}~\bibnamefont
  {Noronha}},\ }\bibfield  {title} {\bibinfo {title} {Causality of the
  einstein-israel-stewart theory with bulk viscosity},\ }\href
  {https://doi.org/10.1103/PhysRevLett.122.221602} {\bibfield  {journal}
  {\bibinfo  {journal} {Phys. Rev. Lett.}\ }\textbf {\bibinfo {volume} {122}},\
  \bibinfo {pages} {221602} (\bibinfo {year} {2019})},\ \Eprint
  {https://arxiv.org/abs/1901.06701} {arXiv:1901.06701 [gr-qc]} \BibitemShut
  {NoStop}%
\bibitem [{\citenamefont {Yang}\ \emph
  {et~al.}(2024{\natexlab{a}})\citenamefont {Yang}, \citenamefont {Hippert},
  \citenamefont {Speranza},\ and\ \citenamefont {Noronha}}]{Yang:2023ogo}%
  \BibitemOpen
  \bibfield  {author} {\bibinfo {author} {\bibfnamefont {Y.}~\bibnamefont
  {Yang}}, \bibinfo {author} {\bibfnamefont {M.}~\bibnamefont {Hippert}},
  \bibinfo {author} {\bibfnamefont {E.}~\bibnamefont {Speranza}},\ and\
  \bibinfo {author} {\bibfnamefont {J.}~\bibnamefont {Noronha}},\ }\bibfield
  {title} {\bibinfo {title} {Far-from-equilibrium bulk-viscous transport
  coefficients in neutron star mergers},\ }\href
  {https://doi.org/10.1103/PhysRevC.109.015805} {\bibfield  {journal} {\bibinfo
   {journal} {Phys. Rev. C}\ }\textbf {\bibinfo {volume} {109}},\ \bibinfo
  {pages} {015805} (\bibinfo {year} {2024}{\natexlab{a}})},\ \Eprint
  {https://arxiv.org/abs/2309.01864} {arXiv:2309.01864 [nucl-th]} \BibitemShut
  {NoStop}%
\bibitem [{\citenamefont {Yang}\ \emph
  {et~al.}(2024{\natexlab{b}})\citenamefont {Yang}, \citenamefont {Hippert},
  \citenamefont {Speranza},\ and\ \citenamefont {Noronha}}]{Yang:2024dsv}%
  \BibitemOpen
  \bibfield  {author} {\bibinfo {author} {\bibfnamefont {Y.}~\bibnamefont
  {Yang}}, \bibinfo {author} {\bibfnamefont {M.}~\bibnamefont {Hippert}},
  \bibinfo {author} {\bibfnamefont {E.}~\bibnamefont {Speranza}},\ and\
  \bibinfo {author} {\bibfnamefont {J.}~\bibnamefont {Noronha}},\ }\bibfield
  {title} {\bibinfo {title} {Second-order transport coefficients in neutron
  star mergers},\ }in\ \href@noop {} {\emph {\bibinfo {booktitle} {Workshop for
  Young Scientists on the Physics of Ultra-relativistic Nucleus-Nucleus
  Collisions}}}\ (\bibinfo {year} {2024})\ \Eprint
  {https://arxiv.org/abs/2402.06878} {arXiv:2402.06878 [nucl-th]} \BibitemShut
  {NoStop}%
\bibitem [{\citenamefont {Yang}\ \emph
  {et~al.}(2024{\natexlab{c}})\citenamefont {Yang}, \citenamefont {Hippert},
  \citenamefont {Speranza},\ and\ \citenamefont {Noronha}}]{Yang:2023ozd}%
  \BibitemOpen
  \bibfield  {author} {\bibinfo {author} {\bibfnamefont {Y.}~\bibnamefont
  {Yang}}, \bibinfo {author} {\bibfnamefont {M.}~\bibnamefont {Hippert}},
  \bibinfo {author} {\bibfnamefont {E.}~\bibnamefont {Speranza}},\ and\
  \bibinfo {author} {\bibfnamefont {J.}~\bibnamefont {Noronha}},\ }\bibfield
  {title} {\bibinfo {title} {Bulk viscosity transport coefficients in neutron
  star mergers},\ }\href {https://doi.org/10.1051/epjconf/202429603003}
  {\bibfield  {journal} {\bibinfo  {journal} {EPJ Web Conf.}\ }\textbf
  {\bibinfo {volume} {296}},\ \bibinfo {pages} {03003} (\bibinfo {year}
  {2024}{\natexlab{c}})},\ \Eprint {https://arxiv.org/abs/2312.09364}
  {arXiv:2312.09364 [nucl-th]} \BibitemShut {NoStop}%
\bibitem [{\citenamefont {Yang}\ \emph {et~al.}(2025)\citenamefont {Yang},
  \citenamefont {Hippert}, \citenamefont {Speranza},\ and\ \citenamefont
  {Noronha}}]{Yang:2025yoo}%
  \BibitemOpen
  \bibfield  {author} {\bibinfo {author} {\bibfnamefont {Y.}~\bibnamefont
  {Yang}}, \bibinfo {author} {\bibfnamefont {M.}~\bibnamefont {Hippert}},
  \bibinfo {author} {\bibfnamefont {E.}~\bibnamefont {Speranza}},\ and\
  \bibinfo {author} {\bibfnamefont {J.}~\bibnamefont {Noronha}},\ }\href@noop
  {} {\bibinfo {title} {Symmetry energy dependence of the bulk viscosity of
  nuclear matter}} (\bibinfo {year} {2025}),\ \Eprint
  {https://arxiv.org/abs/2504.07805} {arXiv:2504.07805 [nucl-th]} \BibitemShut
  {NoStop}%
\bibitem [{\citenamefont {Harutyunyan}\ and\ \citenamefont
  {Sedrakian}(2023)}]{Harutyunyan:2023nvt}%
  \BibitemOpen
  \bibfield  {author} {\bibinfo {author} {\bibfnamefont {A.}~\bibnamefont
  {Harutyunyan}}\ and\ \bibinfo {author} {\bibfnamefont {A.}~\bibnamefont
  {Sedrakian}},\ }\bibfield  {title} {\bibinfo {title} {Phenomenological
  relativistic second-order hydrodynamics for multiflavor fluids},\ }\href
  {https://doi.org/10.3390/sym15020494} {\bibfield  {journal} {\bibinfo
  {journal} {Symmetry}\ }\textbf {\bibinfo {volume} {15}},\ \bibinfo {pages}
  {494} (\bibinfo {year} {2023})},\ \Eprint {https://arxiv.org/abs/2302.09596}
  {arXiv:2302.09596 [nucl-th]} \BibitemShut {NoStop}%
\bibitem [{\citenamefont {Findley}\ \emph {et~al.}(1976)\citenamefont
  {Findley}, \citenamefont {Lai},\ and\ \citenamefont {Onaran}}]{viscobook}%
  \BibitemOpen
  \bibfield  {author} {\bibinfo {author} {\bibfnamefont {W.~N.}\ \bibnamefont
  {Findley}}, \bibinfo {author} {\bibfnamefont {J.~S.}\ \bibnamefont {Lai}},\
  and\ \bibinfo {author} {\bibfnamefont {K.}~\bibnamefont {Onaran}},\
  }\href@noop {} {\emph {\bibinfo {title} {Creep and Relaxation of Nonlinear
  Viscoelastic Materials}}},\ North-Holland Series in Applied Mathematics and
  Mechanics\ (\bibinfo  {publisher} {Dover Publications},\ \bibinfo {year}
  {1976})\BibitemShut {NoStop}%
\bibitem [{\citenamefont {Gavassino}(2023)}]{Gavassino:2023eoz}%
  \BibitemOpen
  \bibfield  {author} {\bibinfo {author} {\bibfnamefont {L.}~\bibnamefont
  {Gavassino}},\ }\bibfield  {title} {\bibinfo {title} {Relativistic bulk
  viscous fluids of burgers type and their presence in neutron stars},\ }\href
  {https://doi.org/10.1088/1361-6382/ace587} {\bibfield  {journal} {\bibinfo
  {journal} {Class. Quant. Grav.}\ }\textbf {\bibinfo {volume} {40}},\ \bibinfo
  {pages} {165008} (\bibinfo {year} {2023})},\ \Eprint
  {https://arxiv.org/abs/2304.05455} {arXiv:2304.05455 [nucl-th]} \BibitemShut
  {NoStop}%
\bibitem [{\citenamefont {Gorda}\ and\ \citenamefont
  {S\"appi}(2022)}]{Gorda:2021gha}%
  \BibitemOpen
  \bibfield  {author} {\bibinfo {author} {\bibfnamefont {T.}~\bibnamefont
  {Gorda}}\ and\ \bibinfo {author} {\bibfnamefont {S.}~\bibnamefont
  {S\"appi}},\ }\bibfield  {title} {\bibinfo {title} {Cool quark matter with
  perturbative quark masses},\ }\href
  {https://doi.org/10.1103/PhysRevD.105.114005} {\bibfield  {journal} {\bibinfo
   {journal} {Phys. Rev. D}\ }\textbf {\bibinfo {volume} {105}},\ \bibinfo
  {pages} {114005} (\bibinfo {year} {2022})},\ \Eprint
  {https://arxiv.org/abs/2112.11472} {arXiv:2112.11472 [hep-ph]} \BibitemShut
  {NoStop}%
\bibitem [{\citenamefont {Gorda}\ \emph {et~al.}(2023)\citenamefont {Gorda},
  \citenamefont {Paatelainen}, \citenamefont {S\"appi},\ and\ \citenamefont
  {Sepp\"anen}}]{Gorda:2023mkk}%
  \BibitemOpen
  \bibfield  {author} {\bibinfo {author} {\bibfnamefont {T.}~\bibnamefont
  {Gorda}}, \bibinfo {author} {\bibfnamefont {R.}~\bibnamefont {Paatelainen}},
  \bibinfo {author} {\bibfnamefont {S.}~\bibnamefont {S\"appi}},\ and\ \bibinfo
  {author} {\bibfnamefont {K.}~\bibnamefont {Sepp\"anen}},\ }\bibfield  {title}
  {\bibinfo {title} {Equation of state of cold quark matter to $o(\alpha_s^3
  \ln \alpha_s)$},\ }\href {https://doi.org/10.1103/PhysRevLett.131.181902}
  {\bibfield  {journal} {\bibinfo  {journal} {Phys. Rev. Lett.}\ }\textbf
  {\bibinfo {volume} {131}},\ \bibinfo {pages} {181902} (\bibinfo {year}
  {2023})},\ \Eprint {https://arxiv.org/abs/2307.08734} {arXiv:2307.08734
  [hep-ph]} \BibitemShut {NoStop}%
\bibitem [{\citenamefont {Kurkela}\ and\ \citenamefont
  {Vuorinen}(2016)}]{Kurkela:2016was}%
  \BibitemOpen
  \bibfield  {author} {\bibinfo {author} {\bibfnamefont {A.}~\bibnamefont
  {Kurkela}}\ and\ \bibinfo {author} {\bibfnamefont {A.}~\bibnamefont
  {Vuorinen}},\ }\bibfield  {title} {\bibinfo {title} {Cool quark matter},\
  }\href {https://doi.org/10.1103/PhysRevLett.117.042501} {\bibfield  {journal}
  {\bibinfo  {journal} {Phys. Rev. Lett.}\ }\textbf {\bibinfo {volume} {117}},\
  \bibinfo {pages} {042501} (\bibinfo {year} {2016})},\ \Eprint
  {https://arxiv.org/abs/1603.00750} {arXiv:1603.00750 [hep-ph]} \BibitemShut
  {NoStop}%
\bibitem [{\citenamefont {Kurkela}\ \emph {et~al.}(2010)\citenamefont
  {Kurkela}, \citenamefont {Romatschke},\ and\ \citenamefont
  {Vuorinen}}]{Kurkela:2009gj}%
  \BibitemOpen
  \bibfield  {author} {\bibinfo {author} {\bibfnamefont {A.}~\bibnamefont
  {Kurkela}}, \bibinfo {author} {\bibfnamefont {P.}~\bibnamefont
  {Romatschke}},\ and\ \bibinfo {author} {\bibfnamefont {A.}~\bibnamefont
  {Vuorinen}},\ }\bibfield  {title} {\bibinfo {title} {Cold quark matter},\
  }\href {https://doi.org/10.1103/PhysRevD.81.105021} {\bibfield  {journal}
  {\bibinfo  {journal} {Phys. Rev. D}\ }\textbf {\bibinfo {volume} {81}},\
  \bibinfo {pages} {105021} (\bibinfo {year} {2010})},\ \Eprint
  {https://arxiv.org/abs/0912.1856} {arXiv:0912.1856 [hep-ph]} \BibitemShut
  {NoStop}%
\bibitem [{\citenamefont {Fraga}\ and\ \citenamefont
  {Romatschke}(2005)}]{Fraga:2004gz}%
  \BibitemOpen
  \bibfield  {author} {\bibinfo {author} {\bibfnamefont {E.~S.}\ \bibnamefont
  {Fraga}}\ and\ \bibinfo {author} {\bibfnamefont {P.}~\bibnamefont
  {Romatschke}},\ }\bibfield  {title} {\bibinfo {title} {The role of quark mass
  in cold and dense perturbative qcd},\ }\href
  {https://doi.org/10.1103/PhysRevD.71.105014} {\bibfield  {journal} {\bibinfo
  {journal} {Phys. Rev. D}\ }\textbf {\bibinfo {volume} {71}},\ \bibinfo
  {pages} {105014} (\bibinfo {year} {2005})},\ \Eprint
  {https://arxiv.org/abs/hep-ph/0412298} {arXiv:hep-ph/0412298} \BibitemShut
  {NoStop}%
\bibitem [{\citenamefont {Laine}\ and\ \citenamefont
  {Schroder}(2006)}]{Laine:2006cp}%
  \BibitemOpen
  \bibfield  {author} {\bibinfo {author} {\bibfnamefont {M.}~\bibnamefont
  {Laine}}\ and\ \bibinfo {author} {\bibfnamefont {Y.}~\bibnamefont
  {Schroder}},\ }\bibfield  {title} {\bibinfo {title} {Quark mass thresholds in
  qcd thermodynamics},\ }\href {https://doi.org/10.1103/PhysRevD.73.085009}
  {\bibfield  {journal} {\bibinfo  {journal} {Phys. Rev. D}\ }\textbf {\bibinfo
  {volume} {73}},\ \bibinfo {pages} {085009} (\bibinfo {year} {2006})},\
  \Eprint {https://arxiv.org/abs/hep-ph/0603048} {arXiv:hep-ph/0603048}
  \BibitemShut {NoStop}%
\bibitem [{\citenamefont {Torres}\ and\ \citenamefont
  {Menezes}(2013)}]{Torres_2013}%
  \BibitemOpen
  \bibfield  {author} {\bibinfo {author} {\bibfnamefont {J.~R.}\ \bibnamefont
  {Torres}}\ and\ \bibinfo {author} {\bibfnamefont {D.~P.}\ \bibnamefont
  {Menezes}},\ }\bibfield  {title} {\bibinfo {title} {Quark matter equation of
  state and stellar properties},\ }\href
  {https://doi.org/10.1209/0295-5075/101/42003} {\bibfield  {journal} {\bibinfo
   {journal} {EPL}\ }\textbf {\bibinfo {volume} {101}},\ \bibinfo {pages}
  {42003} (\bibinfo {year} {2013})},\ \Eprint {https://arxiv.org/abs/1210.2350}
  {arXiv:1210.2350 [nucl-th]} \BibitemShut {NoStop}%
\bibitem [{\citenamefont {Lopes}\ \emph {et~al.}(2021)\citenamefont {Lopes},
  \citenamefont {Biesdorf},\ and\ \citenamefont {Menezes}}]{Lopes_2021}%
  \BibitemOpen
  \bibfield  {author} {\bibinfo {author} {\bibfnamefont {L.~L.}\ \bibnamefont
  {Lopes}}, \bibinfo {author} {\bibfnamefont {C.}~\bibnamefont {Biesdorf}},\
  and\ \bibinfo {author} {\bibfnamefont {D.~{\'e}.~P.}\ \bibnamefont
  {Menezes}},\ }\bibfield  {title} {\bibinfo {title} {Modified mit bag
  models{\textemdash}part i: Thermodynamic consistency, stability windows and
  symmetry group},\ }\href {https://doi.org/10.1088/1402-4896/abef34}
  {\bibfield  {journal} {\bibinfo  {journal} {Phys. Scripta}\ }\textbf
  {\bibinfo {volume} {96}},\ \bibinfo {pages} {065303} (\bibinfo {year}
  {2021})},\ \Eprint {https://arxiv.org/abs/2005.13136} {arXiv:2005.13136
  [hep-ph]} \BibitemShut {NoStop}%
\bibitem [{\citenamefont {Iwamoto}(1982)}]{Iwamoto:1982zz}%
  \BibitemOpen
  \bibfield  {author} {\bibinfo {author} {\bibfnamefont {N.}~\bibnamefont
  {Iwamoto}},\ }\bibfield  {title} {\bibinfo {title} {Neutrino emissivities and
  mean free paths of degenerate quark matter},\ }\href
  {https://doi.org/https://doi.org/10.1016/0003-4916(82)90271-8} {\bibfield
  {journal} {\bibinfo  {journal} {Ann. Phys. (N.Y.)}\ }\textbf {\bibinfo
  {volume} {141}},\ \bibinfo {pages} {1} (\bibinfo {year} {1982})}\BibitemShut
  {NoStop}%
\bibitem [{\citenamefont {Madsen}(1993)}]{Madsen:1993xx}%
  \BibitemOpen
  \bibfield  {author} {\bibinfo {author} {\bibfnamefont {J.}~\bibnamefont
  {Madsen}},\ }\bibfield  {title} {\bibinfo {title} {Rate of the weak reaction
  $s+u\ensuremath{\rightarrow}u+d$ in quark matter},\ }\href
  {https://doi.org/10.1103/PhysRevD.47.325} {\bibfield  {journal} {\bibinfo
  {journal} {Phys. Rev. D}\ }\textbf {\bibinfo {volume} {47}},\ \bibinfo
  {pages} {325} (\bibinfo {year} {1993})}\BibitemShut {NoStop}%
\bibitem [{\citenamefont {Heiselberg}(1992)}]{Heiselberg_1992}%
  \BibitemOpen
  \bibfield  {author} {\bibinfo {author} {\bibfnamefont {H.}~\bibnamefont
  {Heiselberg}},\ }\bibfield  {title} {\bibinfo {title} {The weak conversion
  rate in quark matter},\ }\href {https://doi.org/10.1088/0031-8949/46/6/002}
  {\bibfield  {journal} {\bibinfo  {journal} {Phys. Scr.}\ }\textbf {\bibinfo
  {volume} {46}},\ \bibinfo {pages} {485} (\bibinfo {year} {1992})}\BibitemShut
  {NoStop}%
\bibitem [{\citenamefont {Heiselberg}\ \emph {et~al.}(1991)\citenamefont
  {Heiselberg}, \citenamefont {Baym},\ and\ \citenamefont
  {Pethick}}]{Heiselberg:1991px}%
  \BibitemOpen
  \bibfield  {author} {\bibinfo {author} {\bibfnamefont {H.}~\bibnamefont
  {Heiselberg}}, \bibinfo {author} {\bibfnamefont {G.}~\bibnamefont {Baym}},\
  and\ \bibinfo {author} {\bibfnamefont {C.~J.}\ \bibnamefont {Pethick}},\
  }\bibfield  {title} {\bibinfo {title} {Burning of strange quark matter and
  transport properties of qcd plasmas},\ }\href
  {https://doi.org/10.1016/0920-5632(91)90313-4} {\bibfield  {journal}
  {\bibinfo  {journal} {Nucl. Phys. B Proc. Suppl.}\ }\textbf {\bibinfo
  {volume} {24}},\ \bibinfo {pages} {144} (\bibinfo {year} {1991})}\BibitemShut
  {NoStop}%
\bibitem [{\citenamefont {Koch}(1991)}]{Koch:1991qh}%
  \BibitemOpen
  \bibfield  {author} {\bibinfo {author} {\bibfnamefont {P.}~\bibnamefont
  {Koch}},\ }\bibfield  {title} {\bibinfo {title} {Weak decays of small
  metastable strangelets},\ }\href
  {https://doi.org/10.1016/0920-5632(91)90333-A} {\bibfield  {journal}
  {\bibinfo  {journal} {Nucl. Phys. B Proc. Suppl.}\ }\textbf {\bibinfo
  {volume} {24}},\ \bibinfo {pages} {255} (\bibinfo {year} {1991})}\BibitemShut
  {NoStop}%
\bibitem [{\citenamefont {Iwamoto}(1980)}]{Iwamoto:1980eb}%
  \BibitemOpen
  \bibfield  {author} {\bibinfo {author} {\bibfnamefont {N.}~\bibnamefont
  {Iwamoto}},\ }\bibfield  {title} {\bibinfo {title} {Quark beta decay and the
  cooling of neutron stars},\ }\href
  {https://doi.org/10.1103/PhysRevLett.44.1637} {\bibfield  {journal} {\bibinfo
   {journal} {Phys. Rev. Lett.}\ }\textbf {\bibinfo {volume} {44}},\ \bibinfo
  {pages} {1637} (\bibinfo {year} {1980})}\BibitemShut {NoStop}%
\bibitem [{\citenamefont {Anand}\ \emph {et~al.}(1997)\citenamefont {Anand},
  \citenamefont {Goyal}, \citenamefont {Gupta},\ and\ \citenamefont
  {Singh}}]{Anand2009}%
  \BibitemOpen
  \bibfield  {author} {\bibinfo {author} {\bibfnamefont {J.}~\bibnamefont
  {Anand}}, \bibinfo {author} {\bibfnamefont {A.}~\bibnamefont {Goyal}},
  \bibinfo {author} {\bibfnamefont {V.}~\bibnamefont {Gupta}},\ and\ \bibinfo
  {author} {\bibfnamefont {S.}~\bibnamefont {Singh}},\ }\bibfield  {title}
  {\bibinfo {title} {Burning of two-flavor quark matter into strange matter in
  neutron stars and in supernova cores},\ }\href
  {https://doi.org/10.1086/304063} {\bibfield  {journal} {\bibinfo  {journal}
  {Astrophys. J.}\ }\textbf {\bibinfo {volume} {481}},\ \bibinfo {pages} {954}
  (\bibinfo {year} {1997})}\BibitemShut {NoStop}%
\bibitem [{\citenamefont {Schwenzer}(2012)}]{schwenzer2012longrange}%
  \BibitemOpen
  \bibfield  {author} {\bibinfo {author} {\bibfnamefont {K.}~\bibnamefont
  {Schwenzer}},\ }\href@noop {} {\bibinfo {title} {How long-range interactions
  tune the damping in compact stars}} (\bibinfo {year} {2012}),\ \Eprint
  {https://arxiv.org/abs/1212.5242} {arXiv:1212.5242} \BibitemShut {NoStop}%
\bibitem [{\citenamefont {Navas}\ \emph {et~al.}(2024)\citenamefont {Navas}
  \emph {et~al.}}]{ParticleDataGroup:2024cfk}%
  \BibitemOpen
  \bibfield  {author} {\bibinfo {author} {\bibfnamefont {S.}~\bibnamefont
  {Navas}} \emph {et~al.} (\bibinfo {collaboration} {Particle Data Group}),\
  }\bibfield  {title} {\bibinfo {title} {Review of particle physics},\ }\href
  {https://doi.org/10.1103/PhysRevD.110.030001} {\bibfield  {journal} {\bibinfo
   {journal} {Phys. Rev. D}\ }\textbf {\bibinfo {volume} {110}},\ \bibinfo
  {pages} {030001} (\bibinfo {year} {2024})}\BibitemShut {NoStop}%
\bibitem [{\citenamefont {Fraga}\ \emph {et~al.}(2001)\citenamefont {Fraga},
  \citenamefont {Pisarski},\ and\ \citenamefont
  {Schaffner-Bielich}}]{Fraga:2001id}%
  \BibitemOpen
  \bibfield  {author} {\bibinfo {author} {\bibfnamefont {E.~S.}\ \bibnamefont
  {Fraga}}, \bibinfo {author} {\bibfnamefont {R.~D.}\ \bibnamefont
  {Pisarski}},\ and\ \bibinfo {author} {\bibfnamefont {J.}~\bibnamefont
  {Schaffner-Bielich}},\ }\bibfield  {title} {\bibinfo {title} {Small, dense
  quark stars from perturbative qcd},\ }\href
  {https://doi.org/10.1103/PhysRevD.63.121702} {\bibfield  {journal} {\bibinfo
  {journal} {Phys. Rev. D}\ }\textbf {\bibinfo {volume} {63}},\ \bibinfo
  {pages} {121702} (\bibinfo {year} {2001})},\ \Eprint
  {https://arxiv.org/abs/hep-ph/0101143} {arXiv:hep-ph/0101143} \BibitemShut
  {NoStop}%
\bibitem [{\citenamefont {Alford}\ \emph
  {et~al.}(2024{\natexlab{b}})\citenamefont {Alford}, \citenamefont
  {Harutyunyan}, \citenamefont {Sedrakian},\ and\ \citenamefont
  {Tsiopelas}}]{Alford:2024tyj}%
  \BibitemOpen
  \bibfield  {author} {\bibinfo {author} {\bibfnamefont {M.}~\bibnamefont
  {Alford}}, \bibinfo {author} {\bibfnamefont {A.}~\bibnamefont {Harutyunyan}},
  \bibinfo {author} {\bibfnamefont {A.}~\bibnamefont {Sedrakian}},\ and\
  \bibinfo {author} {\bibfnamefont {S.}~\bibnamefont {Tsiopelas}},\ }\bibfield
  {title} {\bibinfo {title} {Bulk viscosity of two-color superconducting quark
  matter in neutron star mergers},\ }\href
  {https://doi.org/10.1103/PhysRevD.110.L061303} {\bibfield  {journal}
  {\bibinfo  {journal} {Phys. Rev. D}\ }\textbf {\bibinfo {volume} {110}},\
  \bibinfo {pages} {L061303} (\bibinfo {year} {2024}{\natexlab{b}})},\ \Eprint
  {https://arxiv.org/abs/2407.12493} {arXiv:2407.12493 [nucl-th]} \BibitemShut
  {NoStop}%
\bibitem [{\citenamefont {Alford}\ \emph {et~al.}(2025)\citenamefont {Alford},
  \citenamefont {Harutyunyan}, \citenamefont {Sedrakian},\ and\ \citenamefont
  {Tsiopelas}}]{Alford:2025tbp}%
  \BibitemOpen
  \bibfield  {author} {\bibinfo {author} {\bibfnamefont {M.}~\bibnamefont
  {Alford}}, \bibinfo {author} {\bibfnamefont {A.}~\bibnamefont {Harutyunyan}},
  \bibinfo {author} {\bibfnamefont {A.}~\bibnamefont {Sedrakian}},\ and\
  \bibinfo {author} {\bibfnamefont {S.}~\bibnamefont {Tsiopelas}},\ }\href@noop
  {} {\bibinfo {title} {Bulk viscosity of two-flavor color superconducting
  quark matter in neutron star mergers}} (\bibinfo {year} {2025}),\ \Eprint
  {https://arxiv.org/abs/2506.08144} {arXiv:2506.08144 [nucl-th]} \BibitemShut
  {NoStop}%
\bibitem [{\citenamefont {Mannarelli}\ and\ \citenamefont
  {Manuel}(2010)}]{Mannarelli:2009ia}%
  \BibitemOpen
  \bibfield  {author} {\bibinfo {author} {\bibfnamefont {M.}~\bibnamefont
  {Mannarelli}}\ and\ \bibinfo {author} {\bibfnamefont {C.}~\bibnamefont
  {Manuel}},\ }\bibfield  {title} {\bibinfo {title} {Bulk viscosities of a cold
  relativistic superfluid: Color-flavor locked quark matter},\ }\href
  {https://doi.org/10.1103/PhysRevD.81.043002} {\bibfield  {journal} {\bibinfo
  {journal} {Phys. Rev. D}\ }\textbf {\bibinfo {volume} {81}},\ \bibinfo
  {pages} {043002} (\bibinfo {year} {2010})},\ \Eprint
  {https://arxiv.org/abs/0909.4486} {arXiv:0909.4486 [hep-ph]} \BibitemShut
  {NoStop}%
\bibitem [{\citenamefont {Bierkandt}\ and\ \citenamefont
  {Manuel}(2011)}]{Bierkandt:2011zp}%
  \BibitemOpen
  \bibfield  {author} {\bibinfo {author} {\bibfnamefont {R.}~\bibnamefont
  {Bierkandt}}\ and\ \bibinfo {author} {\bibfnamefont {C.}~\bibnamefont
  {Manuel}},\ }\bibfield  {title} {\bibinfo {title} {Bulk viscosity
  coefficients due to phonons and kaons in superfluid color-flavor locked quark
  matter},\ }\href {https://doi.org/10.1103/PhysRevD.84.023004} {\bibfield
  {journal} {\bibinfo  {journal} {Phys. Rev. D}\ }\textbf {\bibinfo {volume}
  {84}},\ \bibinfo {pages} {023004} (\bibinfo {year} {2011})},\ \Eprint
  {https://arxiv.org/abs/1104.5624} {arXiv:1104.5624 [hep-ph]} \BibitemShut
  {NoStop}%
\bibitem [{\citenamefont {Heiselberg}\ and\ \citenamefont
  {Pethick}(1993)}]{Heiselberg:1993cr}%
  \BibitemOpen
  \bibfield  {author} {\bibinfo {author} {\bibfnamefont {H.}~\bibnamefont
  {Heiselberg}}\ and\ \bibinfo {author} {\bibfnamefont {C.~J.}\ \bibnamefont
  {Pethick}},\ }\bibfield  {title} {\bibinfo {title} {Transport and relaxation
  in degenerate quark plasmas},\ }\href
  {https://doi.org/10.1103/PhysRevD.48.2916} {\bibfield  {journal} {\bibinfo
  {journal} {Phys. Rev. D}\ }\textbf {\bibinfo {volume} {48}},\ \bibinfo
  {pages} {2916} (\bibinfo {year} {1993})}\BibitemShut {NoStop}%
\bibitem [{dat(2025)}]{data}%
  \BibitemOpen
  \href {https://doi.org/10.5281/zenodo.15848379} {\bibinfo {title} {Plaintext
  data containing the derivatives of the pressure—10.5281/zenodo.15848379}}
  (\bibinfo {year} {2025})\BibitemShut {NoStop}%
\end{thebibliography}%

\end{document}